\documentclass[runningheads]{llncs}
\usepackage{graphicx}

\usepackage{amssymb}

\usepackage{hyperref}
\usepackage{breakurl}
\usepackage{color,ifthen}
\usepackage{xcolor}

\usepackage{soul}

\usepackage[inline]{enumitem}
\usepackage{colortbl}
\usepackage{array}

\usepackage{commentingCommands}

\newcommand{\cubeName}{UP Cube}
\newcommand{\cubeNameFull}{Usable Privacy Cube}

\newcounter{contextexamp}[section] 
\renewcommand{\thecontextexamp}{C.\arabic{contextexamp}}

\newenvironment{contextexample}[3][]{
\refstepcounter{contextexamp}
\par\vspace{1ex}\noindent%
\textbf{Context example \thecontextexamp\ (#1)}
\noindent\ignorespaces
\begin{quote}
\begin{itshape}
\par\noindent ``#3''%
\hspace{1ex}[#2]
\end{itshape}
\end{quote} 
}{
\hfill\qed
\vspace{1ex}
} 

\newcounter{goals}
\renewcommand{\thegoals}{UPG.\arabic{goals}}
\newcommand{\pgoal}[3]{\refstepcounter{goals}
\par\vspace{1ex}\noindent%
\textbf{\thegoals\ifthenelse{\equal{#1}{}}{\hspace{0ex}}{\ (#1)\hspace{1ex}}}%
\begin{itshape}
#3%
\hspace{1ex}[#2]
\end{itshape}
\vspace{1ex}
}
\newcommand{\refgoal}[1]{\ref{#1}}

\newcounter{contextQuestion}
\renewcommand{\thecontextQuestion}{CQ.\arabic{contextQuestion}}
\newenvironment{cquestion}[1][]{
\refstepcounter{contextQuestion}
\par\vspace{1ex}\noindent%
\textbf{\thecontextQuestion\hspace{1ex}}\ifthenelse{\equal{#1}{}}{\textbf{.}}{\textit{#1}}%
\par\vspace{0.5ex}
}{
\hfill\qed\vspace{1ex}
} 

\newcounter{usabilityCriteria}
\renewcommand{\theusabilityCriteria}{UPC.\arabic{usabilityCriteria}}
\newcounter{usabilitySubCriteria}[usabilityCriteria]
\renewcommand{\theusabilitySubCriteria}{UPC.\arabic{usabilityCriteria}.\arabic{usabilitySubCriteria}}
\newcommand{\uCriteria}[1]{\refstepcounter{usabilityCriteria}
\par\vspace{1ex}\noindent%
\textbf{\theusabilityCriteria\ }%
#1%
}
\newcommand{\uSubCriteria}[1]{\refstepcounter{usabilitySubCriteria}
\par\vspace{1ex}\indent%
\textbf{\theusabilitySubCriteria\ }%
#1%
\vspace{1ex}
}
\newcommand{\uSubCriteriaCount}{\refstepcounter{usabilitySubCriteria}
\textbf{\theusabilitySubCriteria\ }%
}
\newcommand{\refuc}[1]{\ref{#1}}
\newcommand{\longLabCriteriaGoalBased}[1]{\textit{[Based on goal \refgoal{#1}]\allowbreak}}
\newcommand{\shortLabCriteriaGoalBased}[1]{\textit{[\refgoal{#1}]\allowbreak}}
\newcommand{\longLabCriteriaType}[1]{\textit{[Type of criteria: #1]\allowbreak}}
\newcommand{\shortLabCriteriaType}[1]{\textit{[Type: #1]\allowbreak}}
\newcommand{\longLabCriteriaOutcomeUse}[2]{\textit{[#1\ifthenelse{\equal{#2}{}}{}{:#2}]\allowbreak}}
\newcommand{\longEffectiveness}{Effectiveness}
\newcommand{\shortEffectiveness}{Es}
\newcommand{\longEfficiency}{Efficiency}
\newcommand{\shortEfficiency}{Ey}
\newcommand{\longTEFM}{Time used, Human effort expanded, Financial resources expanded, Materials expanded}
\newcommand{\shortTEFM}{TEFM}
\newcommand{\longSatisfaction}{Satisfaction}
\newcommand{\shortSatisfaction}{S}
\newcommand{\longLabCriteriaMeasure}[1]{\textit{[Measure:#1]\allowbreak}}
\newcommand{\shortLabCriteriaMeasure}[1]{\textit{[#1]\allowbreak}}
\newcommand{\longObjective}{Objective}
\newcommand{\shortObjective}{O}
\newcommand{\longPerceived}{Perceived}
\newcommand{\shortPerceived}{P}

\begin{document}
\onlyShortPaper{
\title{
Making GDPR Usable: A Model to Support Usability Evaluations of Privacy
\thanks{A long version of this paper is available as \cite{theTR}.}
\thanks{
We would like to thank the anonymous reviewers for helping improve the paper.
}
}
}
\longPaper{
\title{
Making GDPR Usable: A Model to Support Usability Evaluations of Privacy
\thanks{
This is the long version supporting the paper published as \cite{Johansen2020} in the book edited by M.~Friedewald, M.~\"{O}nen, E.~Lievens, S.~Krenn, and S.~Fricker, titled \emph{``Privacy and Identity Management. Data for Better Living: AI and Privacy''}, volume 576 of series IFIP Advances in Information and Communication Technology, pp. 275--291, published by Springer International Publishing (2020).
\url{https://doi.org/10.1007/978-3-030-42504-3_18}
}
}
}
\titlerunning{A Model to Support Usability Evaluations of Privacy}
%
\author{Johanna Johansen\thanks{The first author was partially supported by the project \href{https://www.iotsec.no/}{IoTSec} -- Security in IoT for Smart Grids, with nr.\ 248113. Thanks go also to Josef Noll for introducing me to the topic of privacy evaluations and labeling.}\inst{1}\orcidID{0000-0003-4908-9045} 
\and Simone Fischer-H\"{u}bner\inst{2}
}
\authorrunning{J.~Johansen \& S.~Fischer-H\"{u}bner}
%
\institute{Department of Informatics, University of Oslo.\\
\email{johanna@johansenresearch.info}
\and 
Department of Mathematics and Computer Science, Karlstad University.
\email{simone.fischer-huebner@kau.se}
}
\maketitle              

\begin{abstract}

We introduce a new model for evaluating privacy that builds on the criteria proposed by the EuroPriSe certification scheme by adding usability criteria.
Our model is visually represented through a cube, called Usable Privacy Cube (or UP Cube), where each of its three axes of variability captures, respectively: rights of the data subjects, privacy principles, and \textit{usable privacy criteria}.
We slightly reorganize the criteria of EuroPriSe to fit with the UP Cube model, i.e., we show how EuroPriSe can be viewed as a combination of only \textit{rights} and \textit{principles}, forming the two axes at the basis of our UP Cube.
In this way we also want to bring out two perspectives on privacy:
that of the data subjects and, respectively, that of the controllers/processors. 
We define usable privacy criteria based on usability goals that we have extracted from the whole text of the 
General Data Protection Regulation.
The criteria are designed to produce measurements of the level of usability with which the goals are reached. 
Precisely, we measure effectiveness, efficiency, and satisfaction, considering both the objective and the perceived usability outcomes, producing measures of accuracy and completeness, of resource utilization (e.g., time, effort, financial), and measures resulting from satisfaction scales.
In the long run, the UP Cube is meant to be the model behind a new certification methodology capable of evaluating the \textit{usability of privacy}, to the benefit of common users. 
For industries, considering also the usability of privacy would allow for greater business differentiation, beyond GDPR compliance. %

\end{abstract}

\section{Introduction}

\longPaper{
\subsection{Motivations}
}

\longPaper{
Privacy is a human right and an essential prerequisite for protecting fundamental human values, such as dignity and autonomy. However, the goal of protecting the privacy of the individuals and public has proved to be challenging to achieve. Currently, people have difficulty in understanding how their privacy is affected by the current practices in the technological world and are often unaware of its value and the implications of its loss. With our work we intend to build on the recent developments in the data protection law and support the business actors to invest into adopting privacy protecting measures. We offer a way to quantify the level of data protection and its usability in technological products and services. Displaying the achieved level of privacy protection, can be used by businesses to compete and differentiate themselves on the market. In addition, the measurements we produce can be translated into visual labels that can inform the users, at their respective level of understanding and interest, about how well a respective product respects their privacy. In this section we give an overview of the current state of privacy protection in technological products and society at large. This is to argue how our work could benefit the society and why the present context is propitious for such work.

With the boom of electronic commerce and the pervasive IoT (Internet of Things) and web services, people are producing enormous amounts of electronic data that are collected by various actors under very imbalanced privacy agreements (i.e., signing Terms of Services on a ``take-it-or-leave-it'' manner). The reasons are multiple; 
for one, it is difficult (if not impossible) for a normal person to know exactly what data her online behavior (or IoT device) produces. 
Then, much of this data is of private nature, but it is difficult to understand which and in what situation. 
Lastly, there are numerous and powerful algorithms (for search, machine-learning, etc.) that can make new (some unthinkable) inferences out of seemingly non-private pieces of data; and even more dangerous when multiple data sources are combined.
} 

\longPaper{
Digital privacy has eluded people up to the point that many have given up the hope to have both privacy and access to digital services; recognized even in the General Data Protection Regulation (GDPR)%
\footnote{GDPR -- General Data Protection Regulation from European Union \cite{EU2016GDPR}.} 
Recital 9 as ``...widespread public perception that there are significant risks to the protection of natural persons, in particular with regard to online privacy''.
One of the reasons, however, is that privacy is a complex concept. It has different personal and contextual connotations and larger social, ethical, legal and political implications that are difficult to grasp by laypeople \cite{ackerman2005privacy}. In the presence of technology, digital privacy becomes even more complicated to understand. For example, there is a constant tension between how much privacy a person can have (i.e., how much of her activity she can keep only to herself) 
and how much authorities are allowed to see in order to ensure safety in the society (e.g., how many and where should public surveillance cameras be placed?). Another uncertainty surrounding privacy comes from the difficulty modern people have in separating their public and private lives 
(e.g., 
use of social media both with coworkers and friends; 
should (or not) data from browsing the Internet be shared with Internet companies or search engines, and when or for what functionality in return?).

When designing for usability, we also need to consider that the main goal of a person when using a piece of technology or a technological system is to fulfill the task or need that the product was intended for, e.g., to buy train tickets from a ticket machine. The primary goal of the train ticket buyer is not to check how well the machine protects her privacy, but to reach a certain destination. The buyer might even be in a hurry, so that it catches the train that leaves in 10 minutes. There will be no time left to read the privacy terms. And even if the buyer has time to read them, in case she does not agree with the terms, she still has to take the train to reach her destination. There might be a buss alternative, but the privacy terms of the buss company might be even more invasive \cite{karat2005usability}.

The complexity of the privacy concept as such and of digital data and technology, make it difficult for one to evaluate the privacy properties of a specific piece of technology (e.g., web service, \onlyShortPaper{Internet of Things (}IoT\onlyShortPaper{)} product, or communication device). The difficulty is not only for average people, but also for regulators to check compliance, and for developers to be able to provide privacy-aware digital services/products/systems.%
\footnote{Note that system/product/service are used interchangeably throughout the paper.}  
Indeed, there are multiple concepts involved in digital privacy, like data sharing (which for normal business practices nowadays can form a highly intricate network of relationships), ownership and control of data, accountability or transparency (both towards the regulators as well as the users). Many of the privacy concepts are even a challenge by themselves, when it comes to their evaluation, since they are difficult to measure or to present/explain.

To solve the intricacies of privacy around the use of personal data in technologies, collaboration between specialists from different fields -- such as computer science, interaction design, human ergonomics, law, and cognitive science -- is needed.
When a common understanding has been reached this should be translated into a form that can be grasped by regular people with limited time and level of expertise in fields such as technology or law. The work of \cite{nissim2017bridging} is an example of such initiative meant to bridge the computer science and legal approaches to privacy.%
\footnote{For this work the authors have received the 2019 PET Award for Outstanding Research in Privacy Enhancing Technologies.}
(See more such examples in Section~\ref{sec_related_works}.)

People having difficulties in evaluating the implications of their choices and behavior in respect to privacy can be taken advantage of by the companies that have business interests in harvesting their data. Moreover, powerful and influential business actors, by using media channels, seek to further misinform people about the value of privacy and encourage them to give up their rights related to privacy and personal data protection.

One type of misconception about privacy that businesses try to spread is that ``only the wrongdoers have something to hide and that the people showing different faces in different situations lack integrity'' \cite[Chapter 10. Privacy]{schneier2015data}. Some renown comments that have shaped the public opinion in this negative way are: 
\begin{itemize}
\item \textit{``You already have zero-privacy anyway, get over it.'' (former Sun Microsystems CEO Scott McNealy, 1999)} \cite[-- source of the citation]{langheinrich2001privacy}

\item \textit{``If you have something that you don't want anyone to know, maybe you shouldn't be doing it in the first place.'' (Google's CEO Eric Schmidt, 2009)}

\item \textit{``Privacy is no longer a social norm.''}\cite[-- source of the citation]{schneier2015data}

\item \textit{``You have one identity. The days of you having a different image for your work friends or co-workers and for the other people you know are probably coming to an end pretty quickly. Having two identities for yourself is an example of a lack of integrity.'' (Facebook's Mark Zuckerberg, 2010)} \cite[-- source of the citation]{schneier2015data}
\end{itemize}
} 

\longPaper{
The companies might be interested in subverting people's privacy so that they can sell them more of their services and products. The information people produce is used mostly for advertising \cite[Chapter 4. The Business of Surveillance]{schneier2015data}. As most of our modern life is conducted online, we do not have much choice, as we cannot simply stop using mobile phones, read or shop online or use e-mail. The solution is to motivate businesses to compete on being ethical and on protecting their users' privacy. Some ways for businesses to achieve this could be: displaying privacy policies shortly and clearly in product descriptions, adopting labels of the energy consumption type, adopting privacy enhancing technologies (PETs), or/and becoming certified by privacy certifications schemes \cite{papakonstantinou2018introduction}. 

One example that consumer communities can do is to include privacy features in technology reviews (e.g., reviews for smartphones), as one main evaluation criterion, at the same level with, e.g., battery life-time. However, people need to be educated to ask for ``privacy-as-a-feature''. The task of educating the public cannot be left in the hands of media, which might be controlled by businesses, or leave the users to learn/study privacy by themselves -- \textit{``if people need to go to great length to protect their privacy, they won't''} \cite{langheinrich2001privacy} because privacy is a too complex matter for a layperson to be able to handle it alone. People need support and guidance in this respect. This can be done through making privacy-related information usable for the general public and motivate the businesses to adopt such an approach to data privacy.
} 

\longPaper{
\subsection{Making privacy usable}
} 

For explaining the intricacies of privacy, besides research articles and books \cite{fra2018handbook}, there are several legislative texts adopted in different jurisdictions.  \longPaper{GDPR}\onlyShortPaper{The General Data Protection Regulation (GDPR)%
\footnote{GDPR -- General Data Protection Regulation from European Union \cite{EU2016GDPR}.}} in Europe makes a good effort in clarifying many aspects of data privacy, providing the legislative support to enforce better data protection practices on anyone (within its jurisdiction) collecting and processing personal data.%
\longPaper{\footnote{The two concepts \textit{``privacy''} and \textit{``data protection''} are used interchangeably throughout this paper. The right to privacy, as emerged in 1984 in the Universal Declaration of Human Rights, is closely related to the nowadays right to protection of personal data. Though the two rights are related, protecting similar values, the right to personal data is though broader as it applies to processing of all kinds of personal data, beyond data related to privacy.\cite{fra2018handbook}. In addition, we address only the privacy issues concerning digital technology.}}
However, these regulations only specify the requirements on the data controllers 
\longPaper{(i.e., the organizations providing the service)}%
in the form of basic principles, and the rights of the data subjects, but do not make any strict claims about the extent to which a controller (or processor) should go about implementing these requirements so that they are beneficial for the user, and to what degree.

As such, one motivation for usability evaluations of privacy is the fact that usability goals of GDPR, s.a.\ 
``... any information ... and communication ... relating to processing [to be provided] to the data subject in a concise, transparent, intelligible  and easily accessible form, using clear and plain language, ...'' (Article 12 (1) of GDPR),
are left open to the subjective interpretation of both evaluators and controllers. The provisions of GDPR regarding usability are too general and high-level to be suitable for a certification process \cite{kamara2018data}.
To remedy this, we propose a set of criteria thought to produce measurable evaluations of the usability with which privacy goals of data protection are reached. 

For evaluating privacy we take as starting point the methodology developed by EuroPriSe%
\longPaper{\footnote{At the moment of writing the paper, EuroPriSe's criteria catalog has not been approved pursuant to Article 42(5) GDPR and EuroPriSe GmbH has not been accredited as a certification body pursuant to Article 43 GDPR yet. EuroPriSe is dedicated to receiving the approval of its certification criteria and the accreditation as a certification body in accordance with Art. 42 f. GDPR asap.}} \cite{EuroPriSe2017}
that has as purpose to evaluate compliance with GDPR. 
We are guided by the EuroPriSe criteria when eliciting, what we call, \textit{principles} and \textit{rights}, which form the two variability axes at the basis 
of our model,
i.e., which principles are followed and which rights are respected.
However, EuroPriSe does not consider usability, which is the main focus of our work here. As such, one contribution of this paper is to show how to add usability aspects to the existing evaluation criteria of EuroPriSe.

Certification schemes (s.a. EuroPriSe) provide a seal showing compliance with data protection regulations and industry standards. In addition to such a certification, our evaluation measures on a scale how well data protection obligations are respected and how easy it is for a user to understand that.
These measurements can be presented to the user in different ways, e.g., using ``traffic  light'' scales, showing which level of usability has been reached by the privacy of a certain technological product. 
A ``traffic light'' presentation of privacy is recommended by \cite[Chapter 6(235)]{HouseOfLoardsReport2016} as a way to ``foster competition'' and ``show good practice on privacy policies''. 

\longPaper{
The marketing motivation for the providers to adopt such a methodology is that the users tend to choose the product that answers best to their specific and real needs for privacy. Only the lack of alternatives in the market today explains why the data subjects still accept detrimental privacy conditions that would rather fit the interest and attitudes of other type of stakeholders (controllers and processors) \cite{schneier2015data}. The metaphor of the ``dancing bear'' of Alan Cooper \cite{cooper2004inmates} illustrates well this situation. 

Creating alternative products or features is a way for the businesses to differentiate themselves. Now a new way of differentiation is the level of privacy protection offered, beyond the minimum required by GDPR, as well as how usable this is. 
Certification schemes such as EuroPriSe will give a product the seal of GDPR compliance \cite{cavoukian2018privacy}, but as the GDPR compliance is mandatory, all businesses will seek to conform.
Beyond this, our methodology would facilitate the differentiation between services by considering usability aspects when implementing measures for protecting privacy. 
Usability is known as a market differentiator, e.g., the ISO 9241-11:2018 standard \cite{ISO9241-11:2018} asserts that designing for usability helps with marketing of a product and with offering the user better customized choices. 
} 

Traditionally, usability is a quality related to the use of a product. In our case, we are not interested in the usability of a product per se, but only in those aspects of a product that concern privacy. Our conceptualization of usable privacy is based on the definition of usability as presented in the ISO 9241-11:2018\onlyShortPaper{ \cite{ISO9241-11:2018}},
which we adapt to include privacy as follows:
\onlyShortPaper{\vspace{-1ex}}
\begin{quote}
\emph{Usable privacy} refers to the extent to which a product or a service protects the privacy of the users in an efficient, effective and satisfactory way by taking into consideration the particular characteristics of the users, goals, tasks, resources, and the technical, physical, social, cultural, and organizational environments in which the product/service is used. 
\end{quote}

Our long term goal is to create a methodology to support service providers to make the privacy of their products more usable. The \cubeNameFull\ (\cubeName) described in Section \ref{sec_UPCube} and the usable privacy criteria introduced in Section \ref{sec_Criteria} are the first building blocks of the methodology we are aiming for. They are meant as tools, for both usability engineering experts and certification bodies, to evaluate if a product was designed to respect and protect the privacy of its users in an usable way. Once privacy measures and privacy enhancing technologies are integrated into the design of a product, it still remains to find out if (and how much or to what extent) those measures empower and respect the rights of their particular user as intended. In Human-Computer Interaction (HCI) this is determined based on user testing and usability evaluations. The criteria we propose presume the use of such established HCI methods for usability evaluations (e.g., \cite{Dumas1999}).

\longPaper{
Furthermore, the usable privacy criteria that we propose for evaluating how efficient or effective a product is in protecting privacy require the technology providers to take into consideration the context of use, including the characteristics and needs of different types of users.
We adopt the definition of \textit{context of use} proposed by ISO 9241-11:2018: 

\begin{quote}
``[context of use] comprises a combination of users, goals, tasks, resources, and the technical, physical and social, cultural and organizational environments in which a system, product or service is used.''                                                                                                                                                                                                        \end{quote}
} 

The legislation does not directly refer to usability goals and context of use as known in the ergonomics/human factors or human-centered design. However, requirements as the one in the Recital (39) of GDPR asking for the information addressed to the data subject to be ``easily accessible and easy to understand'' are categorized in this paper as usability goals, for which we create usable privacy criteria meant to measure effectiveness, efficiency and satisfaction  -- as usability outcomes -- with regard to privacy aspects (we henceforth call these \textit{Usable Privacy criteria}, and abbreviated it as UP criteria). 

After a short digression into related work in Section~\ref{sec_related_works}, we introduce in Section~\ref{sec_UPCube} the \cubeName\ model, which is the main contribution of this work.
We then continue to detail the \cubeName\ in the rest of the paper. Section~\ref{sec_EuroPriSe} presents the EuroPriSe in the new light of the \cubeName, forming the two axes of criteria at its basis.
The third vertical axis of the \cubeName, a genuine contribution of this paper, is formed of the UP criteria detailed in Section~\ref{sec_Criteria}. 
To the best of our knowledge, there is not other work that extends privacy certification schemes with usability criteria.
Section~\ref{sec_Goals} presents usable privacy goals that the criteria are meant to measure.
The \cubeName\ naturally captures Interactions between all the axes, which we talk about in Section~\ref{sec_axesIntersections}.
We conclude in Section~\ref{sec_Conclusion}, presenting also some avenues for further work.

\section{Putting the work into context}\label{sec_related_works}

\vspace{-1ex}\paragraph*{Usable privacy and security.}

The present work can be placed in the research field called \textit{usable privacy and security}, with seminal works s.a.\ \cite{adams1999users,good2003usability,whitten1999johnny,Cranor:2018:SSI:3170427.3185061} and conference series s.a.\ the Symposium On Usable Privacy and Security (SOUPS).
We consider that research on privacy requires, even more than security, an interdisciplinary approach (encompassing the expertise coming from research fields such as Psychology, Law or Human-Computer Interaction). As \cite{ackerman2005privacy} points out, privacy has its meaning rooted in larger cultural and social practices and has political, ethical as well as personal connotations.

\longPaper{
There have been considerable efforts towards including specialist from different areas of research on issues related to privacy. 
Examples of such efforts are the constitution of The Privacy \& Us Innovative Training Network (ITN)%
\footnote{This project has received funding from the European Union’s Horizon 2020 research and innovation program under the Marie Sk\l{}odowska-Curie grant agreement No 675730, within the Marie Sk\l{}odowska-Curie Innovative Training Networks (ITN-ETN) framework;
https://privacyus.eu/}
or the organization of the IFIP Summer School on Privacy and Identity Management%
\footnote{https://www.ifip-summerschool.org/}.
%

Other examples of cross-disciplinary research efforts come from the automation of privacy agreements (or Terms of Services -- ToS) where the goal is to presented ToS in an accessible way to the general user. Notable contributions in this regard are the endeavors of the LeDA network%
\footnote{The Legal Design Alliance (LeDA) is formed of lawyers, designers, technologists, academics, and other professionals who are committed to making the legal system more human-centered and effective, through the use of design. https://www.legaldesignalliance.org/},
The Usable Privacy Policy Project%
\footnote{The Usable Privacy Policy Project, https://usableprivacy.org/. Visit also https://explore.usableprivacy.org/ to navigate privacy policy annotations extracted by both humans and machine learning techniques.} 
or the CLAUDETTE project%
\footnote{CLAUDETTE (automated CLAUse DETectEr), http://claudette.eui.eu/about/index.html. See also their tool http://www.claudette.eu/gdpr/}. 
}

Regarding the relation between security and privacy, in this paper we consider security as one integral aspect of privacy, where privacy implies security but not the other way around. We consider such a clarification necessary, as we have seen a tendency in the general public to equalize the meanings of the two terms in favor of security. In computer science, privacy research has been closely intertwined with security research, reflected e.g.\ in the contents and the structure of the book \cite{Cranor2018Security}. However, in this paper, we favor the term ``usable privacy'', as it includes by default security, which is in accordance with
the data protection legislation, where security (integrity and confidentiality) is specified as one of the several principles to abide by in order to assure the privacy of users' data. 

\vspace{-1ex}\paragraph*{Human-Computer Interaction.}

Having the goal to evaluate the usability of privacy in technological systems and products, makes our work part of the larger HCI research on privacy \cite{ackerman2005privacy,karat2012privacy,karat2005usability,patrick2003human}. Following the classifications made by Iachello and Hong in their review \cite{iachello2007end}, we approach privacy from a ``data protection'' perspective by extracting usability related goals from the GDPR. 
A similar approach is taken in
\onlyShortPaper{\cite{patrick2003human}}%
\longPaper{\cite{patrick2003privacy,patrick2003human}},
which translates legislative clauses of the Directive 95/46/EC (now replaced by GDPR) into interaction implications and interface specifications. 
\longPaper{
Similarly, \cite{langheinrich2001privacy} develops principles for guiding system design based on fair information practices found in the US Privacy Act of 1974 and the EU Directive 95/46/EC.  
}
\longPaper{%
The model we propose integrates well with a user-centered design where HCI methods are applied to elicit requirements based on understanding the users, their needs and the context of use. 
}

For evaluating how well a product meets privacy requirements, context of use variables s.a.\ user capabilities, tasks, the field where the technology is going to be deployed (e.g., healthcare, industrial facilities), should be defined. 
\longPaper{Preferably these definitions should be established in the requirements phase of a product's lifecycle, but definitely these would be defined and considered when running the privacy evaluation based on the UP criteria that we propose here.}
We thus adopt the ergonomic approach 
from ISO 9241-11:2018
where \emph{usability is always considered in a specified context of use}, since the usability to be applied to a certain technology can be significantly different for varied combinations of users, goals, tasks and their respective contexts. 

\longPaper{
ISO/IEC29100:2011 gives a good example of how the context of use is decisive for establishing if a certain type of information can be used to identify a natural person \cite[4.4.2 Other distinguishing characteristics, p.7]{ISO/IEC29100:2011}: 
\begin{quote}
``The last name of a person is insufficient to identify a person at a global scale, but might be enough to identify that person at a company level.''
\end{quote}
} 

\longPaper{
We also make a distinction between user experience goals and usability goals, focusing in this paper on the latter \cite{preece2015interaction}. 
User experience goals are concerned with how users, as individuals, are perceiving a product. As the nature of our work is to find criteria that can be generalized to groups or types of people, to be measurable, and part of an evaluation, usability goals will be more appropriate for such a function. This difference is clearly stressed by the ISO 9241-11:2018 standard, which states that usability typically deals with goals sheared by a user group, while user experience has more emphasis on individual goals.
} 

\section{The Usable Privacy Cube model}\label{sec_UPCube}

\begin{figure}[t]
\centering
\includegraphics[width=0.95\textwidth]{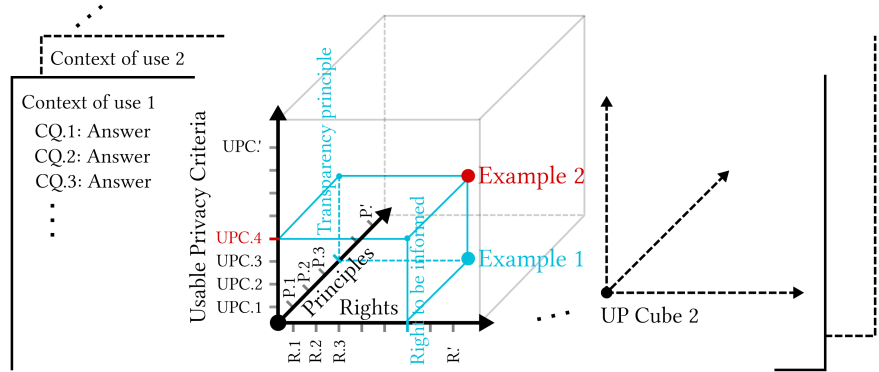}
\onlyShortPaper{\vspace{-2ex}}\caption{A generic version of the cube with the three axes of variability: data protection principles, the rights of the data subjects, and usable privacy criteria.}\onlyShortPaper{\vspace{-2ex}}
\label{fig1}
\end{figure}

We devise a model for organizing the criteria to use in privacy evaluations and measurements, and represent it as a cube with three axes of variability (see Fig.~\ref{fig1}), which we call the \cubeNameFull\ (\cubeName).
The two axes found at the base of the \cubeName\ are composed of the existing EuroPriSe criteria, which we slightly reorganize in the Section~\ref{sec_EuroPriSe} to fit in one of the two categories: data protection principles or rights of the data subjects.

We want to emphasize \textit{two perspectives on privacy} that the \cubeName\ represents (hence our restructuring of the EuroPriSe criteria): the perspective of the controllers and of the data subjects. The controllers are thus given an overview of the principles that they are obliged to follow, whereas the data subjects are offered an overview of their rights.   

The \cubeName\ allows to visualize interactions between the axes, made easier by our separation of the criteria into the three categories. Each such intersection has its specifics and could be studied in itself; we identify a few exemplary points of intersection between the axes in Section~\ref{sec_axesIntersections}.

\begin{example}\label{ex_intersectionPointTransInfo}
The intersection between the transparency principle and the right to be informed is identified in Article 12 of GDPR. The controllers are obliged to provide the data subject information that should be concise, transparent, intelligible and in easily accessible form, using clear and plain language. 
\end{example}

The third vertical axis of the cube is composed of our UP criteria, presented in 
Section~\ref{sec_Criteria}. The UP criteria are determined based on usable privacy goals and are evaluated considering the context of use by following the guidelines in the ISO 9241-11:2018 standard. \longPaper{Each of these criteria has several subcriteria intended for measuring the usability level by using different methods and respective tools from HCI, depending on what the criterion asks to be measured \cite{preece2015interaction}.} 
Interactions exist also with this third axis.


\begin{example}\label{ex_intersectionPointTransInfoUsability}
For the case presented in Example~\ref{ex_intersectionPointTransInfo}, in order to establish how easily accessible or clear the information is, we must measure the level of efficiency, effectiveness and satisfaction in a specific context of use. Efficiency implies measuring the time and effort spent by a specific user for finding the information needed and for understanding it. Effectiveness measures the completeness with which a goal was achieved. In this case we would like to know how much of the needed information was the specific user able to access and understand. At the same time, what a certain type of user perceives as intelligible information, might be perceived by another as difficult to comprehend. Establishing the perceived characteristics of information is an activity categorized under the satisfaction usability outcome.  
\end{example}

The \cubeName\ also brings the idea of \textit{orderings} on each axis, hence the arrows. 
Such orderings are important for several reasons, e.g., 
UP criteria can be ordered based on ``how little effort is required to evaluate it compared to how much overall evaluation outcome it entails'' or ``covers most technologies''.
Usual for certification methods is to use a decision tree order to capture the impact of each criterion (e.g., choosing the most discriminating first), thus which to prioritize in the evaluation.

Judging from practice, one is inclined to think that an ordering is not always possible to find as some principles are equally important, therefore the orders are not necessarily strict. Moreover, one can even see one principle as more important than another only in some industry or context, whereas in a different industry the same two principles would be ordered the other way, therefore one may think that the orders are only partial (i.e., not total).
However, in a specific cube (i.e., used in a specific methodology by a specific authority for privacy usability evaluations in a specific industry and context) there must always be an ordering in which the criteria should be applied. One can always generate a strict and total order from a partial order by just taking a random decision on ordering two criteria when no reasonable order exists. For example, one can any time pick as default order the one arising from the textual placement of the criteria in the data protection legislation texts (maybe considering content from articles as more general than content from recitals), or in the EuroPriSe (or the regulator/company) catalogs. What is certain is that each use case or industry has its specific requirements from which a meaningful ordering would be created.
%

Forming a specific \cubeName, i.e., deciding on the precise details of each criteria on the three axes and the orderings, is to some degree dependent on the specific context of use for the respective product to be evaluated. Therefore, one can think of \textit{infinitely many cubes}, one for each different context. The criteria will not be different between the cubes, but their scope, depth, and evaluation might be different, depending on the context.

\longPaper{

The context of use as such is not mentioned in GDPR, but the \textit{context of processing} is brought up often \cite[e.g., the recitals 43, 47, 71, 74, 76]{EU2016GDPR}. The context of processing, as defined in the legislation, overlaps and has similar purpose with the notion of \textit{context of use} defined in the ISO standard. 
However, unlike usability engineering/HCI where the context is a general concern, the data protection law requires the context to be considered only in certain special situations, e.g., when evaluating the risks to the rights and freedoms of natural persons.
We go beyond this and consider for each proposed criterion the context and the group of users that the evaluation aims at. 
\workInProgress{ Our methodology thus contains questions useful for eliciting knowledge about the users and the context, presented in the Section~\ref{secContext}.}
} 

\workInProgress{
\longPaper{
\section{Context of Use for Privacy}\label{secContext}

For evaluating how well a product meets requirements, variables such as the user capabilities, tasks, the field where the technology is going to be deployed (e.g., healthcare, industrial facilities), and the context of use should be defined.
Preferably these definitions should be established in the requirements phase of a product's lifecycle, but definitely these would be defined and considered when running the privacy evaluation based on the UP criteria that we propose here.
We thus adopt the ergonomic approach \cite{ISO9241-11:2018} where \emph{usability is always considered in a specified context of use}, since the usability to be applied to a certain technology can be significantly different for different combinations of users, goals, tasks and their respective contexts. 
\begin{example}
ISO/IEC29100:2011 gives a good example of how the context of use is decisive for establishing if a certain type of information can be used to identify a natural person \cite[4.4.2 Other distinguishing characteristics, p.7]{ISO/IEC29100:2011}: 
\begin{quote}
``The last name of a person is insufficient to identify a person at a global scale, but might be enough to identify that person at a company level.''
\end{quote}
\end{example}

In the EuroPriSe only some aspects of the context of use (as understood by the ergonomics and HCI communities) are considered: the technical, organizational and legal framework within which the Target of the Evaluation (ToE) is used \cite{hansen2018schleswig}.
However, the GDPR text requires to consider the \textit{context of processing}, which has several overlapping areas with the context of use. Through the data protection by design principle, the context of processing (i.e., nature, scope, context and purposes of processing) as well as the risks for the rights of natural persons, are required to be consider at the time of determining the means of processing and at the time of processing itself [Article 25. Data protection by design and by default of GDPR]. 

\textbf{The context of processing} is brought up in GDPR in several other cases; the ones below are representative.

\begin{contextexample}[for separate consent\label{cex_seaparateConsent}]
{Recital (43) of GDPR}
{Consent is presumed not to be freely given if it does not allow separate consent to be given to different personal data processing operations despite it being \textbf{appropriate in the individual case}, ...}
The ``individual case'' relates to the specific \textit{needs} that a certain data subject might have in respect to using a technology or service. When these are used only partially then the privacy agreement should allow for consenting to processing of data strictly related to the parts used and not to the service as a whole. 
\end{contextexample}

\begin{contextexample}[for legitimate interest of the controller\label{cex_legitInterest}]
{Recital (47) of GDPR}
{At any rate the existence of a legitimate interest would need careful assessment including whether a data subject can reasonably expect \textbf{at the time and in the context} of the collection of the personal data that processing for that purpose may take place.}
One example from healthcare would be that of an emergency situation when the doctors need to get access to the health records of a person being at the moment unable to consent due to serious injuries, and that needs immediate intervention to be rescued.\jjshort{Here and above specific examples from SCOTT can be used.}
\end{contextexample}

\begin{contextexample}[for legitimate interest of data subject]
{Recital (47) of GDPR}
{The interests and fundamental rights of the data subject could in particular override the interest of the data controller where personal data are processed in \textbf{circumstances} where data subjects do not reasonably expect further processing.}
\end{contextexample}

\begin{contextexample}[for further processing\label{cex_furtherProcessing}]
 {Article 6(4)(b) and Recital (50) of GDPR}
 {In order to ascertain whether a purpose of further processing is compatible with the purpose for which the personal data are initially collected, the controller, ..., should take into account, ... the \textbf{context} in which the personal data have been collected, in particular the reasonable expectations of data subjects based on their \textbf{relationship} with the controller as to their further use;}
The relationship mentioned here belongs to the \textit{category of social or organizational context}.
\end{contextexample}

\begin{contextexample}[for risk evaluation of automated decision-making and profiling\label{cex_riskEval}]
 {Recital (71) of GDPR}
 {In order to ensure fair and transparent processing in respect of the data subject, taking into account the \textbf{specific circumstances and context} in which the personal data are processed, ...}
 The specific circumstances and context are considered here for minimizing the risks of automated decision-making and profiling.
\end{contextexample}

\begin{contextexample}[for measures required to be implemented by the controllers in order to promote and safeguard data protection\label{cex_measuresSafeguardPromoteDataProtection}]
 {Recital (74 )of GDPR}
 {Those measures should take into account the \textbf{nature, scope, context and purposes} of the processing and the risk to the rights and freedoms of natural persons.}
The recital refers to the accountability principle, according to which the controllers and processors are required to implement measures to promote and safeguard data protection in their protection activities, but also to demonstrate compliance with the data protection provisions.
\end{contextexample}

\begin{contextexample}[for risk to the rights and freedoms of the data subject\label{cex_rightsFreedomsDataSubject}]
 {Recital (76) of GDPR}
 {The likelihood and severity of the risk to the rights and freedoms of the data subject should be determined by reference to the nature, scope, \textbf{context} and purposes of the processing.}
\end{contextexample}

\subsection{Context eliciting criteria}

The suggestion of \cite{hong2004privacy} to consider the average cases first and then move to the special cases, as the number of exceptions is endless, applies for the use of our questions too. 
The questions we propose to elicit knowledge about the context of use are the following. 
\jjshort{Write about the fact that the questions are based on what is usually considered as context of use in the field of HCI and cite \cite{preece2015interaction}}

\begin{cquestion}[What data is the product/system processing?]\label{cq_dataType}
Information about the type, volatility, size/amount, persistence, accuracy, and value of the data needs to be captured. 

Based on this criteria it can be established if the data is personal or not. If the data gathered cannot be used to identify a natural person, a privacy evaluation will not be needed at all. If the data is of a sensitive nature, an implicit consent needs to be obtained from the data subject. If the data is of a volatile nature then control mechanisms need to be established to periodically check the accuracy and quality of the collected and stored data.
   
The following questions from EuroPriSe can also be considered for eliciting knowledge about the data processed: 
\textit{(i)} questions from section \textit{``1.1.2 Processed Personal Data''} (pp. 16-17), which are meant to establish if any personal data is processed or if these data constitute special kinds of data or data related to criminal convictions and offenses;
and 
\textit{(ii)} questions from section \textit{``C. Target of Evaluation (ToE)''} (p. 13), such as ``What types of personal data are processed when the product or service is used? Which of these data are primary, which are secondary data?''.  
\end{cquestion}

\begin{cquestion}[Who are the privacy stakeholders of the product/solution?]\label{cq_stakeholders}
 
We use the definition 
of \textit{privacy stakeholders} 
from ISO/IEC 291000:2011 \cite{ISO/IEC29100:2011}: 
\begin{quote}
``[Privacy stakeholders] natural or legal person, public authority, agency or any other body that can affect, be affected by, or perceive themselves to be affected by a decision or activity related to''  data processing.                                                                                                                                                                                                       \end{quote}

Considering the ISO/IEC 291000:2011 definition of stakeholders and translating it into GDPR terminology, we can reformulate the question above as follows:
\begin{quote}
\textit{Who are the data subjects, controllers, processors, recipients, third parties of the product/solution?}
\end{quote}

From EuroPriSe, the following questions are relevant to identify stakeholders: 
\textit{(i)} from section \textit{``C. Target of Evaluation (ToE)''} (pp. 13-14) ``Which groups of data subjects are concerned when the product or service is used (e.g., consumers, employees of corporate customers, employees of the service provider)?''; 
and 
\textit{(ii)} from section \textit{``1.1.3 Controller''} (pp. 17-18) the questions meant to identify the controller of each processing operation.
\end{cquestion}

\begin{cquestion}[Which are the specific characteristics/attitudes/expectations/\allowbreak abilities/\allowbreak be\-haviors/\allowbreak capabilities/needs of the data subjects that need to be considered when designing for privacy?]\label{cq_userTypes}

Further classifications of users are recommended, e.g., based on their attitudes towards privacy and privacy preferences, such as the classification of 
\cite{westin1991harris} or the sub-scales of \cite{smith1996information}. 
Assigning users to categories is an integral part of the requirements analysis phase of the usability engineering lifecycle. Personas are created to represent the key characteristics, skill sets, goals, top tasks, responsibilities, tools, or pain points of a category of users \cite[Chapter 41: User Experience Requirements Analysis within the Usability Engineering Lifecycle]{jacko2012human}. Some common classifications are made based on the user's abilities and skills, educational background or the frequency of use of the respective technology \cite[Chapter 10: Establishing Requirements]{preece2015interaction}. 

ISO/IEC29100:2011 \cite[4.5.4 Other factors, p.7]{ISO/IEC29100:2011} also acknowledges the importance of understanding the expectations and preferences of a natural person in respect to privacy. The capacity of a natural person to evaluate privacy risks depends on how well that person understands the technology s/he uses, her/his past experience or background, and other socio-psychological factors. 
\end{cquestion}

\begin{cquestion}[Which are the characteristics of the data subject's work environment and social context that could influence the privacy related requirements?]\label{cq_characteristicsWorkEnvironment}
\end{cquestion}

\begin{cquestion}[Which are the specific characteristics/needs/restriction of the considered industry or organization in respect to privacy?]\label{cq_characteristicsIndustry}

Examples from the medical field: For assuring a balance between protecting patient information and allowing access by the clinicians in emergency medical situations, the processing should be done only under the responsibility of a professional who is subject to the obligation of professional secrecy. 

One should identify whether privacy standards already exist for a specific industry (e.g., the health care industry has its own privacy standards for disclosure of patient identifiable information or training of healthcare workforce \cite[Chapter 30: Human-Computer Interaction in Health Care]{jacko2012human}).

The question in the section \textit{``C. Target of Evaluation (ToE)''} of EuroPriSe (p. 14) -- What is the area of application of the product or service (e.g., is the product to be used in the medical or in the advertising sector)? -- is relevant as a generic question in this category. 
\end{cquestion}

\begin{cquestion}[What are the functional requirements for the system/product that might have privacy implications?]\label{cq_functionalRequirements}

The functional requirements refer to what a product should do. 
In the example of an Electronic Health Records (EHR) system, this is meant to gather electronically health-related information on an individual that can be created, managed, and consulted by authorized clinicians and staff across several care organizations \cite[Chapter 30: Human-Computer Interaction in Health Care]{jacko2012human}. From this functional description it becomes clear that the system is meant to deal with sensitive personal data. 

For the context of use, \cite[Chapter 10: Establishing Requirements]{preece2015interaction} identifies four aspects of the environment that must be considered: the physical, social, organizational, and technical environment.
\end{cquestion}

\begin{cquestion}[Which aspects of the physical context could influence the privacy of a product/solution?]\label{cq_physicalContext}
 
If the environment is very crowded, an IoT device could unintentionally gather data on other people happening to be in that area. 
ISO/IEC29100:2011 \cite[4.4.6 Unsolicited PII, p.9]{ISO/IEC29100:2011} says that collecting unsolicited data can be reduced by considering privacy safeguarding measures at the time of the design of the system. An example of a safeguarding measure in the case when the employees of an organization work in an open office landscape where people outside of the organization might get access, can be to lock the computer containing sensitive data before leaving for the day or being out of the office for longer periods.
\end{cquestion}

\begin{cquestion}[Which aspects of the social context could influence the privacy of a product/solution?]\label{cq_socialContext}

This criteria is concerned with establishing a good overview of who is collaborating with whom and how their work is coordinated, how they communicate and on which issues; e.g., data might need to be pseudonymized before being shared with other members of the organization, for further processing, that are not authorized to have access to that data.
\end{cquestion}

\begin{cquestion}[Which aspects of the organizational context could influence the privacy of a product/solution?]\label{cq_organizationalContext}

Privacy controls such as legal contracts, management practices, procedures, guidelines, organizational structures need to be put in place to safeguard the data and be properly communicated to all parties involved in the data processing.  
\end{cquestion}

\begin{cquestion}[Which aspects of the technical environment could influence the privacy of a product/solution?]

The type of the technology that is used to gather data, the existing technological limitations, and compatibility problems between applications and technologies need to be known. Langheinrich \cite{langheinrich2001privacy} and Hong and et. al. \cite{hong2004privacy} give an overview of the privacy aspects concerning ubiquitous technologies (e.g., sensors embedded in our environments to gather temperature, light, and noise information can be object to secondary use and be analyzed to deduce information on living patterns of the respective individual). 

The question in the section \textit{``C. Target of Evaluation (ToE)''} of EuroPriSe meant to define what is the Target of Evaluation (p. 13) -- which components of the product, which interfaces, which hardware and software components -- are relevant for this category. 
\end{cquestion}
}
}

\section{EuroPriSe}\label{sec_EuroPriSe}


EuroPriSe originated\jjshort{Rewrite this paragraph, and see not to repeat things below or in footnotes.} from the Schleswig-Holstein Data Protection Seal, which was led by the Schleswig-Holstein Data Protection Authority (DPA) from ca.~2001 until the end of 2013, when it was transferred to the EuroPriSe GmbH company. 
The scheme has a history of eighteen years \cite{hansen2018schleswig} and is one of the oldest privacy and data protection seals based on a law, i.e., the State Data Protection Act of the German federal State Schleswig-Holstein. The role of the seal is to help the vendors of IT products and services to comply with the data protection requirements derived from the applicable law in Europe \cite{balboni2018controversies,cavoukian2018privacy,papakonstantinou2018introduction}. 
\longPaper{EuroPriSe, in collaboration with 
Unabh\"{a}ngiges Landeszentrum f\"{u}r Datenschutz Schleswig-Holstein (ULD),
 received support from EU to establish a trans-European privacy seal. EuroPriSe is now intended to provide EU-wide privacy certifications that assure compliance with European data protection law.}
In addition, the EuroPriSe criteria are already updated to consider the fairly new GDPR. 

We have chosen EuroPriSe as the basis for our \cubeName\ because of its long history, its continuous improvement, strong list of well-developed criteria, being led 
in the past by a DPA%
, and being based on the European data protection legislation. 
EuroPriSe also integrates with widely acknowledged IT security certification methods s.a.~ISO 27000 and the The Standard Data Protection Model
\footnote{Following the requirement for a consistency mechanism set out in the Article 63 of GDPR, the work of the certifications bodies and DPAs in Germany is coordinated and made consistent through \textit{``The Standard Data Protection Model''} (\url{https://www.datenschutz-mv.de/datenschutz/datenschutzmodell/}), issued by the Conference of the Independent Data Protection Authorities of the Bund and the L\"{a}nder \longPaper{(Germany)} on 9-10 November 2016. This document is a good reference for methods and guidance for implementing the data protection principles.}. 

\onlyShortPaper{
\begin{table}[!t]
\begin{tabular}{|p{10.2cm}|>{\centering}p{0.5cm}|>{\centering}p{0.5cm}|>{\centering\arraybackslash}p{0.5cm}|}
    \hline\rowcolor[RGB]{190,235,248}
    \vspace{0.5ex}\textbf{EuroPriSe Criteria:} 
    We list the names of (sub)sections as appearing in the EuroPriSe document  \cite{EuroPriSe2017}, which has two parts, the second being subdivided into four \emph{sets} of criteria, whereas the first contains preliminary issues, from where only section C is relevant for us.
    \vspace{1ex} \  & \rotatebox{-90}{Principles\ } & \rotatebox{-90}{Rights\ } & \rotatebox{-90}{Context\ }\\
    \hline
    C. Target of Evaluation (ToE) & \checkmark  & & \checkmark \\
    \hline
    1.1.1 Processing Operations; Purpose(s) & \checkmark & & \checkmark \\
    \hline
    1.1.2 Processed Personal Data & & & \checkmark \\
    \hline
    1.1.3 Controller & & & \checkmark \\
    \hline
    1.1.4 Transnational Operations & & & \checkmark \\
    \hline
    1.2.1 Data Protection by Design and by Default & \checkmark & & \\
    \hline
    1.2.2 Transparency & \checkmark & & \\
    \hline
    2.1 Legal Basis for the Processing of Personal Data & \checkmark & & \\
    \hline
    2.2 General Requirements & \checkmark & & \\
    \hline
    2.3.1 Data Collection (Information Duties) & & \checkmark & \\
    \hline
    2.3.2 Internal Data Disclosure & \checkmark & \checkmark & \\
    \hline
    2.3.3 Disclosure of Data to Third Parties & \checkmark & \checkmark & \\
    \hline
    2.3.4 Erasure of Data after Cessation of Requirement & & \checkmark & \\
    \hline
    2.4.1 Processing of Data by Joint Controllers & \checkmark & & \\
    \hline
    2.4.2 Processing of Data by a Processor & \checkmark & & \\
    \hline
    2.4.3 Transfer to the Third Countries & \checkmark & & \\
    \hline
    2.4.4 Automated Individual Decisions & \checkmark & & \\
    \hline
    2.4.5 Processing of Personal Data Relating to Children & & & \checkmark \\
    \hline
    2.5 Compliance with General Data Protection Principles & \checkmark & & \\
    \hline
    Set 3: Technical-Organisational Measures & \checkmark & & \\
    \hline
    Set 4: Data Subjects' Rights & & \checkmark & \\
    \hline
\end{tabular}
\vspace{1ex}
\caption{Overview of the the EuroPriSe criteria categorized to fit into our \cubeName\ model, i.e., as the two axes with Principles and Rights, as well as Context of use.}\label{table_euroPriSeCriteria}
\onlyShortPaper{\vspace{-4ex}}
\end{table} 
}
 
The way the criteria are formulated, as questions, also fits with the form of our usable privacy evaluation criteria.
In addition, the existing EuroPriSe evaluation, which is at the basis of our model, assures that the GDPR legal grounds are covered, including data protection principles and duties and data subject rights. The UP criteria evaluations come on top, fine-graining the EuroPriSe evaluation with usability measurements, showing how well the legislation is respected.


Another feature that is relevant for our user-centered approach is that the EuroPriSe criteria catalog has been updated to include the data protection by default paradigm, promoting built-in data protection and privacy-friendly default settings. 
Moreover, EuroPriSe takes into account the technical, organizational and legal framework within which the product or service is operated and asks for considering the requirements of all the parties involved in the system, aiming at strengthening the position of the data subjects. Our work shares with EuroPriSe its high-level goal of making transparent for the general public how companies are managing data protection in their products and services. 

\longPaper{
\subsection{The EuroPriSe criteria in the \cubeName\ structure}\label{subsec_EuroPriSe_evaluations}
} 


\longPaper{}

In order to build on EuroPriSe, we first look into how its methodology fits with our \cubeName\ model. We show how EuroPriSe criteria can be redistributed into one of the two axes at the basis, i.e., as either rights of the data subjects or as privacy principles, or otherwise as a context of use criterion. Table \ref{table_euroPriSeCriteria} gives an overview of this redistribution.
The distinction between principles and rights is inspired by the structure in \cite{fra2018handbook}, where principles and rights represent the core of this handbook. One purpose of the principles, mentioned in \cite{fra2018handbook}, is to serve as the starting point when interpreting the more detailed provisions in the subsequent articles of data protection law. 
The law also requires that these principles should correspond to the rights presented in the articles 12 to 22. This correspondence can be visualized through the intersection between the respective rights and principles axes of the \cubeName. 

In the following subsections we detail the Table \ref{table_euroPriSeCriteria}.

\longPaper{
\subsubsection{Data protection principles.}
EuroPriSe has a dedicated subset of criteria dealing with data protection principles, i.e., ``2.5 Compliance with General Data Protection Principles''. In addition, several other criteria sets from EuroPriSe can be related to data protection principles, as detailed below.

\begin{itemize}
 \item The second part of the section ``C. Target of Evaluation (ToE)'', called ``Regulatory Analysis'', as well as the ``Purpose(s)'' part of the Subset 1.1.1, refer to the \textit{principle of purpose limitation}. This principle requires that the purpose of data processing must be defined before processing is started.
 
 \item The ``1.2.1 Data Protection by Design and by Default'' refers to the \textit{data minimization principle}. This is pointed out in a note introducing the subset.
 
 \item The criterion ``How long are the data retained? Is this no longer than necessary for the purposes concerned?'' of the ``1.2.1.1 Data protection by Design'' refers to the \textit{storage limitation principle}.
 
  \item ``1.2.2 Transparency'' relates to the \textit{transparency of processing principle}. 
  
  \item ``2.1 Legal Basis for the Processing of Personal Data'' refers to the \textit{principle of lawfulness}. Consent of the data subject or another legitimate ground (provided in the data protection legislation) are required as legal basis for processing personal data. This subset expounds as well on aspects that we categorize as belonging to the context of use\workInProgress{ such as \textbf{\ref{cq_dataType}} or \textbf{\ref{cq_characteristicsIndustry}}}.
  
  \item The following sets of criteria are related to the \textit{accountability principle}:
   \begin{itemize}
    \item The subset ``2.2.1 Record of Processing Activities'' details how controllers can facilitate compliance with the accountability requirement through recording processing activities and making them available to the supervisory authority upon request (this is also explained in \cite[3.7. The accountability principle]{fra2018handbook}).
    
    \item The subset ``2.2.2 Designation of a Data Protection Officer'' details how controllers can facilitate compliance with the accountability requirement through designating a data protection officer who is involved in all issues relating to personal data protection (this is also explained in \cite[3.7. The accountability principle]{fra2018handbook}). 
    
    \item The subset ``2.2.4 Data Protection Impact Assessment'' details how controllers can facilitate compliance with the accountability requirement through undertaking data protection impact assessments for types of processing likely to result in high risks to the rights and freedoms of natural persons (this is also explained in \cite[3.7. The accountability principle]{fra2018handbook}).
    
    \item The subset ``2.2.5 Prior consultation'' details how compliance is promoted through prior consultation of the relevant supervisory authority if the impact assessment indicates that processing presents risks that cannot be mitigated (this is also explained in \cite[4.3. Rules on accountability and promoting compliance]{fra2018handbook}).
   \end{itemize}
   
\item The following sets are related to the \textit{data security principle}:
    \begin{itemize}
 
    \item The subset ``2.2.6 Notification of a personal Data Breach'' details how the controller is required to notify the competent supervisory authority without undue delay, in cases where a personal data breach (with risks for rights and freedoms of individuals) takes place. The data subject being concerned needs to be informed as well (this is also explained in \cite[3.6. The data security principle]{fra2018handbook}).
    
    \item The set ``Set 3: Technical-Organisational Measures: Accompanying Measures for Protection of the Data Subject'' follows for the most the \cite[Article 32: Security of processing]{EU2016GDPR}.

    \end{itemize}
 
\end{itemize}

\subsubsection{The rights of the data subjects.}
EuroPriSe has a dedicated subset of criteria dealing with data subjects' rights: ``Set 4: Data Subjects' Rights''. In addition, several other criteria sets from EuroPriSe can be related to rights of the data subjects, as detailed bellow.
\begin{itemize}
 \item ``2.3.1 Data Collection (Information Duties)'' refers to the \textit{right to be informed}, following \cite[Articles 12, 13 and 14]{EU2016GDPR}.
 \item ``2.3.4 Erasure of Data after Cessation of Requirement'' refers to the \textit{right to erasure}, following \cite[Article 17]{EU2016GDPR}.
 \item ``2.4.4 Automated Individual Decisions'' refers to the rights related to \textit{automated individual decision-making}, following \cite[Article 22]{EU2016GDPR}.
\end{itemize}

\subsubsection{Mixed and context of use}

\begin{itemize}
\item ``2.3.2 Internal Data Disclosure'' and ``2.3.3 Disclosure of Data to Third Parties'' refer to a mixture of rights and principles. These subsets have pointers to the GDPR articles they are based on, which are indicators for which rights and principles the sets can be linked to.  Article 5(b) refers to the \textit{principle of purpose limitation}, (c) to the \textit{principle of data limitation}, (f) to the \textit{data security principle}, article 6 to the \textit{lawfulness of processing principle}, while articles 13 and 14 to the \textit{right to be informed}.

\item Section ``C. Target of Evaluation'' and other subsections from 1.1 and from 2.4 are seen as relevant for defining the context of use.

 \item Criteria relevant for defining stakeholder groups\workInProgress{ in \textbf{\ref{cq_stakeholders}}} are found in ``2.2.7 Processing under the Authority of the Controller or Processor'', ``2.4.1 Processing of data by Joint Controllers'', and ``2.4.2 Processing of Data by a Processor'', and regard obligations of the controllers, processors, joint controllers, and third parties. These groups are also mentioned in \cite[2.3. Users of personal data]{fra2018handbook} as users of the data. The criteria set ``2.2.3 Designation of the Representative in the EU'' defines representatives within the EU; these are also mentioned in \cite[Data protection terminology]{fra2018handbook}, in the context of defining the controllers group. When the controller is established outside the EU it needs to appoint a representative within the EU territory.
  
  However, some of the criteria included in these sets could also be related to principles or rights, e.g., the criterion ``Does the processor adhere to an approved code of conduct or an approved certification mechanism?'' (``2.4.2 Processing of data by a Processor'') belongs to the \textit{accountability principle}, which states that the controllers must be able to demonstrate compliance with data protection provisions.

\workInProgress{
\begin{itemize}
 \item Section \textit{``C. Target of Evaluation (ToE)/ ToE Analysis''} helps with defining the object of the evaluation, types of data, groups of data subjects and area of application. As shown in section \ref{secContext}, such type of criteria are used to define the context of use: \textbf{\ref{cq_dataType}},\textbf{ \ref{cq_stakeholders}}, \textbf{\ref{cq_characteristicsIndustry}}. 
 \item  The questions in section \textit{``1.1.1 Processing operations''} are meant to define the operations associated with the use of a product or service, which relate in our case with \textbf{\ref{cq_functionalRequirements}}.
 \item \textit{``1.1.2.2 Special Categories of Personal Data''} and \textit{``1.1.2.3 Personal Data Relating to Criminal Convictions and Offenses''} relate to the \textbf{\ref{cq_dataType}}.
 \item \textit{``1.1.3 Controller''} relates to \textbf{\ref{cq_stakeholders}}.
 \item \textit{``1.1.4 Transnational Operations''} and \textit{``2.4.3 Transfer to the Third Countries''} point to the physical context \textbf{\ref{cq_physicalContext}}, which encompasses in this case a larger territorial view, extending beyond the physical space of a building (office building or private house), to countries or transnational areas where the product might be used or data might be processed.
 \item \textit{``2.4.5 Processing of Personal Data Relating to Children''} refer to \textbf{\ref{cq_dataType}},\textbf{ \ref{cq_stakeholders}}. 
\end{itemize}
}

\end{itemize}

} 

\section{Usable Privacy Goals}\label{sec_Goals}

We identify usable privacy goals (henceforth called \textit{Usable Privacy goals}, and abbreviated as UP goals) that appear in the GDPR text. These guide the work in Section~\ref{sec_Criteria} where we present the UP criteria meant to measure to what extent these goals are being achieved. 
\onlyShortPaper{We give here only some examples of goals, numbered as in the long version \cite{theTR}, where the full list of 30 UP goals can be found.}
The goals are listed in the order they appear in the legislation.
The words \textbf{\textit{emphasized}} in each goal relate to usability.
The chosen words are those that can be interpreted differently based on the context they are used in, and can result in objective and perceived measurements when evaluated in usability tests. These words also capture goals that can be achieved up to certain degrees, and thus can be translated into a level in an evaluation scale.
\longPaper{
A list of the Recitals and Articles from GDPR where the goals were extracted from can be found in full in the Annex A. 
}
In addition to the GDPR, there are more specific data protection laws, such as the proposed ePrivacy Regulation, that have implications for usability, from where one could eventually extract additional usability goals.

\longPaper{
\pgoal{}{Recital (6) of GDPR}{
\label{goal_generic_level}
Ensuring a \textbf{high level of protection} of personal data.
}

\pgoal{}{Recital (7) of GDPR}{
\label{goal_control} 
Natural persons should have \textbf{control} of their own personal data.
}
}

\onlyShortPaper{\setcounter{goals}{2}}
\pgoal{}{Recital (32) of GDPR}{
\label{goal_consent_clearFreelyGivenSpecific}
Consent should be given by a \textbf{clear} affirmative act establishing a \textbf{freely given}, \textbf{specific}, \textbf{informed} and \textbf{unambiguous} indication of the data subject's agreement to the processing of personal data relating to him or her.
}

\longPaper{
\pgoal{}{Recital (32) of GDPR}{\label{goal_consent_clearConcise}
If the data subject's consent is to be given following a request by electronic means, the request must be \textbf{clear}, \textbf{concise} and \textbf{not unnecessarily disruptive} to the use of the service for which it is provided.
}
}  

\longPaper{
\pgoal{}{Recital (39) of GDPR}{\label{goal_info_easilyAccesible}
Any information and communication related to the processing of personal data should be \textbf{easily accessible} and \textbf{easy to understand}.
}
}  

\longPaper{
\pgoal{}{Recital (39) of GDPR}{\label{goal_info_clearPlainLanguage}
Any information and communication related to the processing of personal data should use \textbf{clear and plain language}.
}

\pgoal{}{Recital (39) of GDPR}{\label{goal_risk_awareness}
Make the natural persons \textbf{aware} of risks, rules, safeguards and rights in relation to the processing of personal data. 
}
}  

\onlyShortPaper{\setcounter{goals}{7}}
\pgoal{}{Recital (39) of GDPR}{\label{goal_rights_awareness}
Make the natural persons \textbf{aware} of how to exercise their rights in relation to processing of personal data.
}

\longPaper{
\pgoal{}{Recital (39) of GDPR}{\label{goal_purpose_explicit}
The specific purposes for which personal data are processed should be \textbf{explicit}.
}

\pgoal{}{Recital (39) of GDPR}{\label{goal_purpose_adequate}
The personal data should be \textbf{adequate}, \textbf{relevant} and limited to what is \textbf{necessary} for the purposes for which they are processed.
}

\pgoal{}{Recital (39) of GDPR}{\label{goal_purpose_reasonably}
Personal data should be processed only if the purpose of the processing could not \textbf{reasonably} be fulfilled by other means.
}
}  

\longPaper{
\pgoal{}{Recital (42) of GDPR}{\label{goal_consent_awareFactExtent}
In the context of a written declaration on another matter, safeguards should ensure that the data subject is \textbf{aware} of the fact that and the extent to which consent is given. 
}
}  

\longPaper{
\pgoal{}{Recital (42) of GDPR}{\label{goal_consent_intelligible}
A declaration of consent pre-formulated by the controller should be provided in an \textbf{intelligible} and \textbf{easily accessible} form, using \textbf{clear} and \textbf{plain language} and it should not contain \textbf{unfair terms}. 
}
}  

\longPaper{
\pgoal{}{Recital (42) of GDPR}{\label{goal_consent_genuine}
The data subject should have \textbf{genuine and free choice} in giving the consent.}
}  

\longPaper{
\pgoal{}{Recital (42) of GDPR}{\label{goal_consent_withdraw}
The data subject should be able to refuse or withdraw consent \textbf{without detriment}.
}
}  

\longPaper{
\pgoal{}{Recital (47) of GDPR}{\label{goal_legitimateInterest_controller}
Carefully assess the existence of a legitimate interest of a controller taking into consideration the \textbf{reasonable} expectations of data subjects based on their relationship with the controller.
}

\pgoal{}{Recital (47) of GDPR}{\label{goal_legitimateInterest_dataSubject}
Assess if the interests and fundamental rights of the data subject could override the interest of the controller where personal data are processed in circumstances where data subjects do not \textbf{reasonably} expect further processing. 
}
}  

\onlyShortPaper{\setcounter{goals}{17}}
\pgoal{}{Article 12 (1) and Recital (58) of GDPR}{\label{goal_info_concise}
Any information addressed to the public or to the data subject should be \textbf{concise}, \textbf{easily accessible} and \textbf{easy to understand}.
}

\pgoal{}{Article 12 (1) and Recital (58) of GDPR}{\label{goal_info_clearPlainLanguage_Second}
Any information addressed to the public or to the data subject should use \textbf{clear and plain language}.
}

\longPaper{
\pgoal{}{Recital (58) of GDPR}{\label{goal_info_appropriateVisualization}
Any information addressed to the public or to the data subject should use, when \textbf{appropriate}, \textbf{visualization}.
}
}  

\onlyShortPaper{\setcounter{goals}{20}}
\pgoal{}{Article 12 (7) and Recital (60) of GDPR}{\label{goal_info_easilyVisible}
Provide information of the intended processing in an \textbf{easily visible}, \textbf{intelligible} and \textbf{clearly legible} manner.
}

\longPaper{
\pgoal{}{Article 12 (7) and Recital (60) of GDPR}{\label{goal_info_overviewProcessing}
Provide a \textbf{meaningful} overview of the intended processing.
}
}  

\longPaper{
\pgoal{}{Recital (63) of GDPR}{\label{goal_rights_easilyAccess}
A data subject should have the right of access to personal data which have been collected concerning him or her, and should exercise that right \textbf{easily and at reasonable intervals}, in order to be aware of, and verify, the lawfulness of the processing.
}
}  

\onlyShortPaper{\setcounter{goals}{23}}
\pgoal{}{Recital (100) 
linking to Article 42 of GDPR
}{\label{goal_generic_quicklyAssessDataProtectionLevel}
Allow the data subjects to \textbf{quickly assess} the level of data protection of relevant products and services.%
}

\longPaper{
\pgoal{}{Article 7 (2) of GDPR}{\label{goal_consent_requestClearlyDistinguishable}
The request for consent should be presented in a manner which is \textbf{clearly distinguishable} from the other matters, in an \textbf{intelligible} and \textbf{easily accessible} form, using \textbf{clear and plain language}.
}
}  

\longPaper{
\pgoal{}{Article 7 (3) of GDPR}{\label{goal_consent_easyWithdraw}
It should be \textbf{as easy} to withdraw as to give consent.
}
}  

\longPaper{
\pgoal{}{Article 12 (2) of GDPR}{\label{goal_rights_facilitate}
\textbf{Facilitate} the exercise of the data subjects rights under Articles 15 to 22  -- right of access, right to rectification, right to erasure, right to restriction of processing, right to data portability, right to object and automated individual decision-making.
}
}  

\longPaper{
\pgoal{}{Article 15 (1) (h) of GDPR}{\label{goal_info_meaningfulAutomatedDecisionMaking}
The data subject should obtain from the controller \textbf{meaningful} information about the logic involved, as well as the \textbf{significance} and the \textbf{envisaged consequences} of automated decision-making, including profiling to which s/he is object to.
}
}  

\longPaper{
\pgoal{}{Article 21 (4) of GDPR}{\label{goal_rights_rightToObject}
The right to object should be \textbf{explicitly} brought to the attention of the data subject and should be presented \textbf{clearly} and separately from any other information, at the latest at the time of the first communication with the data subject.
}
}  

\longPaper{
\pgoal{}{Article 22 (1) of GDPR}{\label{goal_rights_automatedProcessing}
The data subject should have the right not to be subject to a decision based solely on automated processing, including profiling, which produces legal effects concerning him or her or similarly \textbf{significantly} affects him or her.
}
}  
 
%

\section{Usable Privacy Criteria}\label{sec_Criteria}

\longPaper{
The criteria presented in this section can be used in an evaluation process for establishing the level of effectiveness, efficiency, and satisfaction with which the goals from Section~\ref{sec_Goals} are reached, wrt.\ a specific context of use.
} 

The proposed criteria are always measurable, which makes the results of a privacy evaluation easier to present visually through the use of a \textit{privacy labeling} scheme.
The use of privacy labels will then fulfill the goal \refgoal{goal_generic_quicklyAssessDataProtectionLevel}. This goal has a special significance from a usability point of view as it reduces considerably the effort spent by the data subject for evaluating privacy, which for most users is not the primary task \cite{ackerman2005privacy} and it gets in the way of buying or using a product or service.

\longPaper{
Evaluating privacy and compliance with GDPR is done by certification bodies, providing seals and marks with the purpose of enhancing consumer trust and promoting transparency and compliance with the data protection regulations. Prior to GDPR, the lack of legal constraints, disconnection from official regulatory oversight, and lack of effective enforcement has resulted in inaccurate, false, or outdated privacy certificates. This made the existing certification schemes to lose their trustworthiness with the users (e.g., see in \cite{edelman2011adverse} criticisms of  TRUSTe -- now known as TrustArc\footnote{\url{https://en.wikipedia.org/wiki/TrustArc\#Criticism\_and\_Controversies}}).
Though the certification is still voluntary, GDPR endorses and facilitates a certification mechanism as a means to demonstrate compliance with data protection provisions. In addition, the existence of a certificate makes the process of choosing processors easier for the controllers too, especially so since GDPR establishes responsibility and liability for any processing carried out on the controller's behalf \cite{kamara2018data}.
}  

\longPaper{
\subsection{General considerations for UP Criteria}
}  

\longPaper{

Some of the goals have a more general purpose, and their achieved levels can be decided based on measurements coming from more specific criteria. In the case of the \longPaper{\refgoal{goal_control}}\onlyShortPaper{\cite[UPG.2]{theTR}}, in order to establish if the data subjects have reached a \textit{high level} of ``control of their own personal data'', the scores from evaluations of e.g.,  \longPaper{\refgoal{goal_consent_genuine}}\onlyShortPaper{\cite[UPG.14]{theTR}} should also be high.

Another example of a general goal is \refgoal{goal_purpose_reasonably}, which related to the data minimization principle. This goal is of special importance for our model since when there is no processing we give automatically the highest score on the evaluation scale.
The criterion in the section \textit{``1.1.2.1 Personal data''} of EuroPriSe: ``Are any personal data processed when the product or service is used?'' can be used to establish if personal data is being processed. This can be complemented by another EuroPriSe criterion, found in the section \textit{``1.2.1.1 Data Protection by design''} (p. 18): ``Is it possible to carry out the processing without the use of identifiable data all together?''. This criterion has the function to encourage the companies -- if possible -- to not process identifiable data at all.

} 

The \longPaper{\refgoal{goal_generic_level}}\onlyShortPaper{\cite[UPG.1]{theTR}} is another general goal, dealing with the protection of personal data in general. We use this goal to exemplify how a criterion should be formulated to consider usability:
\begin{quote}
What is the level of the \textit{usability} of the personal data protection / privacy that the product or service ensures?
\end{quote} 

For being able to establish a level of how usable the privacy protection is, the evaluation needs to produce \textit{measurable} outcomes. The structure that we follow is the one proposed in the ISO 9241-11:2018 where the measures consider both the objective and the perceived outcomes of usability (the UP criteria are labeled accordingly). 
The measurements will produce \textit{counts} or \textit{frequencies} (e.g., how many errors the user does when probed to do certain privacy related tasks) and \textit{continuous data} (e.g., how much time does the user spend on completing a task related to privacy). 
The evaluation based on the UP criteria proposed below will produce three \textit{main categories of measures}: 
\begin{enumerate}
\item measures of accuracy and completeness, 
\item resource utilization (time, effort, financial, and material resources), and 
\item measures resulting from satisfaction scales.
\end{enumerate}
%
In our structuring of the UP criteria we first give a high-level criterion, numbered e.g.~\onlyShortPaper{\refuc{criteria_info_conciseEasilyAccessible}}\longPaper{\refuc{criteria_generic_control}}, which is closely related -- which can also be seen from the wording -- to one of the goals that we identified in Section~\ref{sec_Goals}. 
The score for a main UP criterion is established based on evaluations of more specific UP criteria, called subcriteria. 
The resources used to achieve a criterion, i.e., \textit{time, effort, financial, and material} (which we abbreviate TEFM), should be measured to be able to determine the efficiency with which a specific criterion was reached. In addition, the results from the evaluations should show the level of perception that the data subjects have about their data being protected. The data subjects need to be highly satisfied with the offered privacy protection. The ``high satisfaction'' level is defined based on the user satisfaction evaluation of the respective subcriteria.

\longPaper{
\subsubsection{Notation and organization principles.}

We use here as example the first UP criterion, numbered \refuc{criteria_generic_control}, to explain the organization of the UP criteria and the various notations that we use in the rest of this section.
}  

\begin{figure}[t]
\centering
\includegraphics[width=1\textwidth]{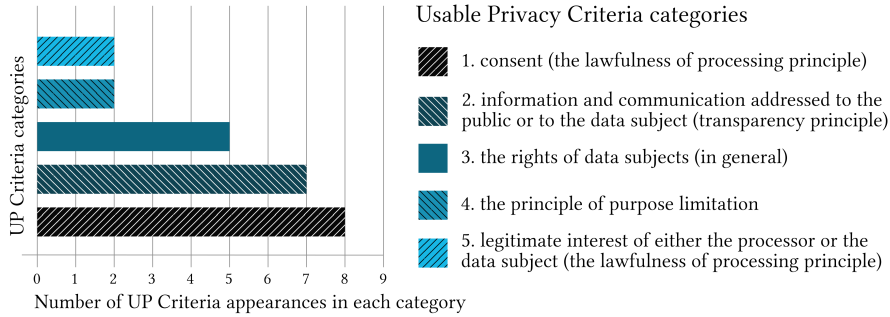}
\caption{An overview of the distribution of usable privacy criteria in each category.}\onlyShortPaper{\vspace{-2ex}}
\label{fig2}
\end{figure}

We categorize the UP criteria based on their area of application from the GDPR text. Figure~\ref{fig2} gives an overview of the number of criteria in each category.
\longPaper{
\begin{enumerate}
 \item Consent (lawful grounds for processing data principle).
 \item Information and communication addressed to the public or to the data subject (transparency principle), 
 \item Rights of the data subjects (rights in general), 
 \item Purpose of processing. 
 \item Legitimate interest of either the processor or the data subject (lawful grounds for processing data principle). 
 \end{enumerate}
 } 
 
\onlyShortPaper{A high-level UP criterion, like \refuc{criteria_info_conciseEasilyAccessible}, is labeled with the goal that it is related to, \refuc{goal_info_concise}.}%
\longPaper{
A high-level UP criterion, like \refuc{criteria_generic_control}, is labeled with the goal that it is related to, e.g.: 

\begin{center}
\longLabCriteriaGoalBased{goal_control}.
\end{center}
Each UP criterion also is \textit{categorized} using a label, e.g.:  
\begin{center}
\longLabCriteriaType{generic}.
\end{center}
\longPaper{
For subsequent UP criteria we use a short version of these labels, which should be self-explanatory, e.g., we label \refuc{criteria_info_conciseEasilyAccessible} by the goal that it aims at \shortLabCriteriaGoalBased{goal_info_concise}.
}  

For each high-level UP criterion we identify several specific UP subcriteria, numbered accordingly, e.g., \refuc{criteria_info_conciseEasilyAccessible_Completeness_access}.
} 
\longPaper{

\paragraph*{Considerations regarding the classification of UP subcriteria.}
}
We then classify each UP subcriterion (e.g., from \refuc{criteria_info_conciseEasilyAccessible_TEFM_access} to \refuc{criteria_info_conciseEasilyAccessible_percieve_understand}) into either effectiveness, efficiency, or satisfaction, and label it accordingly\onlyShortPaper{.}\longPaper{, e.g.: 
\begin{center}
a short version of the label \longLabCriteriaOutcomeUse{\longEffectiveness}{} would be \longLabCriteriaOutcomeUse{\shortEffectiveness}{}; 
\end{center}
similarly \longLabCriteriaOutcomeUse{\shortEfficiency}{} and respectively \longLabCriteriaOutcomeUse{\shortSatisfaction}{}.
} 
\longPaper{

For the first UP criterion \refuc{criteria_generic_control}, we give \textit{objective} and \textit{perceived} subcriteria. 
This is the structure we recommend to be used in a real evaluation. However, with the intention of reducing the complexity of the present paper, for the subsequent UP criteria, we only label the various subcriteria with the respective labels and sublabels of effectiveness, efficiency, and satisfaction, e.g.:  
\begin{center}
\refuc{criteria_generic_controlEs} is labeled with \longLabCriteriaOutcomeUse{\longEffectiveness}{} and \longLabCriteriaMeasure{\longObjective}. 
\end{center}
}
We try to be exhaustive in our UP subcriteria and to give enough questions to cover all major aspects that need to be measured to achieve the respective goal that the high-level UP criterion relates to. The UP subcriteria are labeled with sublabels representing various specific measures of usability for the above three general categories, e.g.:
\onlyShortPaper{\longLabCriteriaOutcomeUse{\longEffectiveness}{Completeness}.}
\longPaper{
\begin{center}
\longLabCriteriaOutcomeUse{\longEffectiveness}{Completeness} or \\
\longLabCriteriaOutcomeUse{\longSatisfaction}{Cognitive responses}.
\end{center}
} 

\longPaper{
\paragraph*{Considerations regarding the context of use.}

The \textbf{specific context} of use needs to be considered for each of our questions. To avoid repetition, we only give one example of how the questions should be formulated so that they relate to the context. This formulation applies to all the questions we propose in this section. 
For the example of \refuc{criteria_generic_control} one would read it:
\begin{itemize}
\item Without context \\ 
\textit{What is the level of control the data subjects have over their data?}

\item With context \\
\textit{What is the level of control the \textbf{specified type of} data subjects have over their data \textbf{in the specified context of use}?} 
\end{itemize}
}  

\subsection{List of Usable Privacy criteria}\label{subsec_CriteriaList}

\onlyShortPaper{
We give few examples of the usable privacy criteria, while the full list of all 24 UP criteria can be found in the long version \cite{theTR}.
} 
Since our criteria are modular (i.e., each high-level criterion is thought independent of the other) and can be ordered based on their importance for different application cases, they could be introduced gradually and selectively. 
It can be that certification bodies (like EuroPriSe) would start to include our UP criteria in their future test catalogs on an article-basis, e.g., a good candidate is Article 12 of GDPR (referring to rights that intersect with the transparency principle) as it contains five UP goals.

\vspace{2ex}
\longPaper{
\uCriteria{\label{criteria_generic_control} What is the level of control that the data subjects have over their data?} \longLabCriteriaGoalBased{goal_control}\longLabCriteriaType{generic}
%

\begin{itemize}
 \item[] \uSubCriteria{\label{criteria_generic_controlEs} To what degree are the data subjects in control of the personal data? \longLabCriteriaOutcomeUse{\longEffectiveness}{}\longLabCriteriaMeasure{\longObjective}}

 \item[] \uSubCriteria{\label{criteria_generic_controlEs2} What is the data subjects' perceived level of control? \longLabCriteriaOutcomeUse{\longEffectiveness}{}\longLabCriteriaMeasure{\allowbreak\longPerceived}}

 \item[] \uSubCriteria{\label{criteria_generic_controlEy} How much [Time / Effort / Financial / Material resources] do the data subjects need to invest in order to have control over the processed data? \longLabCriteriaOutcomeUse{\longEfficiency}{\longTEFM}\longLabCriteriaMeasure{\longObjective}}

 \item[] \uSubCriteria{\label{criteria_generic_controlEy2} How much [Time / Effort / Financial / Material resources] do the data subjects perceive that they need to invest in order to have control over the processed data? \longLabCriteriaOutcomeUse{\longEfficiency}{\longTEFM}\longLabCriteriaMeasure{\longPerceived}}

 \item[] \uSubCriteria{\label{criteria_generic_controlS} How frequent do the data subjects use the data controlling tools put to their disposition?
 \longLabCriteriaOutcomeUse{\longSatisfaction}{}\longLabCriteriaMeasure{\longObjective}}

 \item[] \uSubCriteria{\label{criteria_generic_controlS2} What is the level of satisfaction of the data subjects with the achieved level of control? \longLabCriteriaOutcomeUse{\longSatisfaction}{}\longLabCriteriaMeasure{\longPerceived}}
\end{itemize}

\subsubsection{UP criteria related to information and communication}\ 

}

\onlyShortPaper{\setcounter{usabilityCriteria}{1}}
\uCriteria{\label{criteria_info_conciseEasilyAccessible} Is any information and communication addressed to the public or to the data subjects related to the processing of personal data concise, easily accessible and easy to understand?} \shortLabCriteriaGoalBased{goal_info_concise}\longLabCriteriaType{Information and communication addressed to the public or to the data subjects}
\begin{itemize}
 \item[] How much TEFM do the data subjects need to invest in order to 
 [\uSubCriteriaCount\label{criteria_info_conciseEasilyAccessible_TEFM_access} access, 
 \uSubCriteriaCount\label{criteria_info_conciseEasilyAccessible_TEFM_read} read through, 
 \uSubCriteriaCount\label{criteria_info_conciseEasilyAccessible_TEFM_understand} understand]
 the information? 
 \longLabCriteriaOutcomeUse{\longEfficiency}{\shortTEFM}\shortLabCriteriaMeasure{\shortObjective}%
\vspace{1ex}

\vspace{1ex}
\item[] How much of the information were the data subjects able to 
[\uSubCriteriaCount\label{criteria_info_conciseEasilyAccessible_Completeness_access} access, 
\uSubCriteriaCount\label{criteria_info_conciseEasilyAccessible_Completeness_understand} understand, 
\uSubCriteriaCount\label{criteria_info_conciseEasilyAccessible_Completeness_read} read through]? 
\shortLabCriteriaMeasure{\shortObjective}\longLabCriteriaOutcomeUse{\shortEffectiveness}{Completeness}

\vspace{1ex}
\item[] \uSubCriteria{\label{criteria_info_conciseEasilyAccessible_concise} To what degree the data subjects perceive the information as concise?
\longLabCriteriaOutcomeUse{\longSatisfaction}{Cognitive responses}
\shortLabCriteriaMeasure{\shortPerceived}
}

\vspace{1ex}
\item[] To what degree the data subjects perceive the information as easy to \phantom{abcdef} 
[\uSubCriteriaCount\label{criteria_info_conciseEasilyAccessible_percieve_access} access, 
\uSubCriteriaCount\label{criteria_info_conciseEasilyAccessible_percieve_understand} understand]?
\longLabCriteriaOutcomeUse{\longSatisfaction}{Cognitive responses}\shortLabCriteriaMeasure{\shortPerceived}
\end{itemize}

\begin{remark}
 The subcriteria in \refuc{criteria_info_conciseEasilyAccessible} refer to cognition and understanding, while the subcriteria in \refuc{criteria_info_visibleLegible} refer to visual aspects of the information presented.
\end{remark}

\begin{remark}
In different HCI works one can find different formulations that could seem related to how we formulate the subcriteria, e.g.:
``Can the data subjects make sense of the information at all?''; 
``What is the extent to which the data subjects make sense of the information?''.
However, we intend to measure the proportion of the information that is made sense of. Therefore we use formulations that give a statistically measurable outcome, such as ``How much?'', ``What is the percentage?'', ``What is the degree?''.
\end{remark}

\uCriteria{\label{criteria_info_visibleLegible} Is the information about the intended processing provided in an easily visible, intelligible and clearly legible manner?} \shortLabCriteriaGoalBased{goal_info_easilyVisible}\shortLabCriteriaType{Info}
\begin{itemize}
 \item[] How much TEFM do the data subjects need to invest in order to [\uSubCriteriaCount see/locate, and \uSubCriteriaCount distinguish] the information? 
 \longLabCriteriaOutcomeUse{\shortEfficiency}{\shortTEFM}  
\vspace{1ex}
 
 \item[] How well were the data subjects able to [\uSubCriteriaCount visually locate and \uSubCriteriaCount distinguish] the information?
 \longLabCriteriaOutcomeUse{\shortEffectiveness}{Accuracy}
\vspace{1ex}

 \item[] How much of the information were the data subjects able to [\uSubCriteriaCount visually locate and \uSubCriteriaCount distinguish]? 
 \longLabCriteriaOutcomeUse{\shortEffectiveness}{Completeness}
\vspace{1ex}
 
 \item[] To what degree the data subjects perceive the information as [\uSubCriteriaCount easily visible, \uSubCriteriaCount intelligible, and \uSubCriteriaCount clearly legible]?
 \longLabCriteriaOutcomeUse{\shortSatisfaction}{Cognitive responses}
\end{itemize}

\begin{remark}
Poor visibility can affect the perception of trust, as information that has low visibility can appear to be hidden with a purpose. 
Poor legibility can reflect sloppiness in the way the content is produced, which again can give an impression of lack of professionalism.
Poor visibility and legibility affects the satisfaction of the data subjects and it can cause physical discomfort (e.g., to the eyes, by having to read a text written in a very small font).
\end{remark}

\uCriteria{\label{criteria_info_clearPlainLanguage} Is any information and communication addressed to the public or to the data subjects related to the processing of personal data using clear and plain language?}
\shortLabCriteriaGoalBased{goal_info_clearPlainLanguage_Second}\shortLabCriteriaType{Info}
\begin{itemize} 
 
 \item[] What is the level of [\uSubCriteriaCount clearness and \uSubCriteriaCount plainness] of the language?
 \longLabCriteriaOutcomeUse{\shortEffectiveness}{Accuracy}
 
 \item[] \uSubCriteria{What is the percentage of the data subjects that understand the language?
 \longLabCriteriaOutcomeUse{\shortEffectiveness}{Completeness}
}
 
 \item[] What is the portion of the language considered [\uSubCriteriaCount plain and \uSubCriteriaCount clear]?
 \longLabCriteriaOutcomeUse{\shortEffectiveness}{Completeness}
\vspace{1ex}
 
 \item[] How [\uSubCriteriaCount clear and \uSubCriteriaCount plain] do the data subjects perceive the language to be?
 \longLabCriteriaOutcomeUse{\shortSatisfaction}{Cognitive responses}
\end{itemize}

\longPaper{
\uCriteria{\label{criteria_info_appropriateVisualization}Is the information and communication addressed to the public or to the data subjects using, when appropriate, visualization?} 
\shortLabCriteriaGoalBased{goal_info_appropriateVisualization}\shortLabCriteriaType{Info}
\begin{itemize} 
 \item[] \uSubCriteria{How much of the data subjects' expended TEFM is reduced by the use of visualization?
 \longLabCriteriaOutcomeUse{\shortEfficiency}{\shortTEFM}
}
 
 \item[] \uSubCriteria{How well is the information understood when visualization is used, in comparison to when only text is used? 
 \longLabCriteriaOutcomeUse{\shortEffectiveness}{Accuracy}
}
 
 \item[] \uSubCriteria{How many data subjects consider the use of visualization in the evaluated context of use as appropriate? 
 \longLabCriteriaOutcomeUse{\shortEffectiveness}{Accuracy}\shortLabCriteriaMeasure{\shortPerceived}
}
 
 \item[] \uSubCriteria{What is the percentage of data subjects that understand the information better, when visualization is used?
 \longLabCriteriaOutcomeUse{\shortEffectiveness}{Completeness}
}
 
 \item[] \uSubCriteria{To what degree is the understanding of the information improved by the use of visualization?
 \longLabCriteriaOutcomeUse{\shortEffectiveness}{Completeness}
}
 
 \item[] \uSubCriteria{What is the level of satisfaction of the data subjects when visualization is used?
 \longLabCriteriaOutcomeUse{\shortSatisfaction}{Cognitive responses}
}
 
 \item[] \uSubCriteria{How appropriate do the data subjects perceive the use of visualization to be for the evaluated context of use?
 \longLabCriteriaOutcomeUse{\shortSatisfaction}{Cognitive responses}
}
\end{itemize}

\begin{remark}
 Some of the subcriteria in \refuc{criteria_info_appropriateVisualization} mention the ``understanding of information'' in relation with visualization. However, measurements of other aspects, besides cognitive effort, such as how visualization improves the rate of finding and accessing information, should be evaluated here as well.
\end{remark}
} 

\longPaper{
\uCriteria{\label{criteria_info_overviewProcessing} Are the data subjects provided a meaningful overview of the intended processing?}
\shortLabCriteriaGoalBased{goal_info_overviewProcessing}\shortLabCriteriaType{Info}
\begin{itemize}
 \item[] \uSubCriteria{How much of the data subjects' expended TEFM is reduced by using the provided overview?
 \longLabCriteriaOutcomeUse{\shortEfficiency}{\shortTEFM}
}

\vspace{1ex}
 \item[] What is the percentage of the data subjects that 
[\uSubCriteriaCount use and 
\uSubCriteriaCount understand the content better due to] 
the provided overview?
 \longLabCriteriaOutcomeUse{\shortEffectiveness}{Accuracy}

%

\vspace{1ex}
 \item[] \uSubCriteria{What is the degree of improvement that the overview brings to the understanding of the content by data subjects? 
 \longLabCriteriaOutcomeUse{\shortEffectiveness}{Completeness}
}
 
 \item[] \uSubCriteria{What is the percentage of data subjects able to express the correct and intended meaning of the provided overview, when probed?
 \\ \longLabCriteriaOutcomeUse{\shortEffectiveness}{Completeness}
}
 
 \item[] \uSubCriteria{How meaningful do the data subjects perceive the provided overview?
 \longLabCriteriaOutcomeUse{\shortSatisfaction}{Cognitive responses}
}
\end{itemize}
} 

\longPaper{
\uCriteria{\label{criteria_info_meaningfulAutomatedDecisionMaking} Have the data subjects obtained from the controller meaningful information about $<\!\!<$the logic involved, as well as the significance and the envisaged consequences of automated decision-making, including profiling to which they are object to$>\!\!>$ ?}
\shortLabCriteriaGoalBased{goal_info_meaningfulAutomatedDecisionMaking}\shortLabCriteriaType{Info}

\begin{remark}
 To avoid repetition, in the subordinated subcriteria, we write the above text between angle brackets in the following short form: ${<\!\!<}$LOGIC${>\!\!>}$.
\end{remark}

\begin{itemize}
 \item[] \uSubCriteria{How much TEFM do the data subjects need to invest in order to understand the information about ${<\!\!<}$LOGIC${>\!\!>}$?
 \longLabCriteriaOutcomeUse{\shortEfficiency}{\shortTEFM}
}
 
 \item[] \uSubCriteria{What is the percentage of data subjects being able to express the correct and intended meaning of the information provided -- in respect to ${<\!\!<}$LOGIC${>\!\!>}$ -- when probed?
 \longLabCriteriaOutcomeUse{\shortEffectiveness}{Accuracy}
}

\vspace{1ex}
 \item[] To what degree do the provided information 
[\uSubCriteriaCount affect the choices and actions and 
\uSubCriteriaCount improve the understanding] 
of the data subjects in respect to ${<\!\!<}$LOGIC${>\!\!>}$?
\longLabCriteriaOutcomeUse{\shortEffectiveness}{Accuracy}
resp.\ \longLabCriteriaOutcomeUse{\shortEffectiveness}{Completeness}

%
 
 \item[] \uSubCriteria{How much of the provided information -- in respect to ${<\!\!<}$LOGIC${>\!\!>}$ -- is understood by the data subjects?
 \longLabCriteriaOutcomeUse{\shortEffectiveness}{Completeness}
}
 
 \item[] \uSubCriteria{How meaningful do the data subjects perceive the provided information in respect to ${<\!\!<}$LOGIC${>\!\!>}$?
 \longLabCriteriaOutcomeUse{\shortSatisfaction}{Cognitive responses}
}
\end{itemize}
} 

\longPaper{
\subsubsection{UP criteria related to consent} \ 
}

Several UP goals are found in the consent related provisions. These provisions are evaluated in detail in the EuroPriSe sections \textit{2.1.1.1 Processing on the Basis of Consent} and \textit{2.1.1.2 Processing on the Basis of a Contract}. 
The criteria we generate here are meant to complement the ones in the EuroPriSe through bringing in usability concerns. 
Marc Langheinrich presents several of the problems with how consent can be misused \cite{langheinrich2001privacy}. One of these is the ``take it or leave it'' dualism where the person does not have a real choice and thus getting consent comes very closed to blackmailing. This problem has been ameliorated in the GDPR law by asking the controllers to allow for separate consent for different data processing operations. A usability evaluation could help further by revealing how the data subjects perceive the consenting act, as well as whether the data subjects consider consent a real choice and if the options to consent to some of the processing operations only, are satisfactory. 
\jjshort{TODO: Write about the dark patterns used by companies based on the psychological tendencies in users.}

\vspace{2ex}
\onlyShortPaper{\setcounter{usabilityCriteria}{7}}
\uCriteria{Is consent given by a clear affirmative act establishing a freely given, specific, informed and unambiguous indication of the data subjects' agreement to the processing of personal data relating to them?}
\shortLabCriteriaGoalBased{goal_consent_clearFreelyGivenSpecific}\shortLabCriteriaType{Consent}
\begin{itemize}
 \item[] \uSubCriteria{How much of the consent text do the data subjects understand?
 \longLabCriteriaOutcomeUse{\shortEffectiveness}{Completeness}
}
 
 \item[] \uSubCriteria{How much of the implications of consenting do the data subjects understand?
 \longLabCriteriaOutcomeUse{\shortEffectiveness}{Completeness}
}

\vspace{1ex}
 \item[] To what degree do the data subjects perceive the agreement to be
[\uSubCriteriaCount freely given, 
\uSubCriteriaCount informed, and
\uSubCriteriaCount unambiguous]?
\longLabCriteriaOutcomeUse{\shortSatisfaction}{Cognitive responses}
\end{itemize}
%

\longPaper{
\uCriteria{Are the consents of the data subjects given by following a request by electronic means? If yes, is the request clear, concise and not unnecessarily disruptive to the use of the service for which it is provided?}
\shortLabCriteriaGoalBased{goal_consent_clearConcise}\shortLabCriteriaType{Consent}
\begin{itemize}
 \item[] \uSubCriteria{\label{criteria_consent_electronic_TEFM}How much TEFM do the data subjects need to invest in order to understand the request?
 \longLabCriteriaOutcomeUse{\shortEfficiency}{\shortTEFM}
}

 \item[] \uSubCriteria{How much of the request do the data subjects understand?
 \\\longLabCriteriaOutcomeUse{\shortEffectiveness}{Completeness}
}
 
 \item[] \uSubCriteria{How much of the TEFM needed to fulfill the tasks that the data subjects are currently doing, is being wasted by attending to the request?
 \longLabCriteriaOutcomeUse{\shortEfficiency}{\shortTEFM}
}

 \item[] \uSubCriteria{To what degree do the data subjects perceive the request to be unnecessarily disruptive?
\longLabCriteriaOutcomeUse{\shortSatisfaction}{Cognitive responses}
}
\end{itemize}
} 

\longPaper{
\uCriteria{In the context of a written declaration on another matter, are safeguards ensured so that the data subjects are aware of the fact that, and the extent to which, consent is given?}
\shortLabCriteriaGoalBased{goal_consent_awareFactExtent}\shortLabCriteriaType{Consent}
\begin{itemize}
\item[] \uSubCriteria{What is the percentage of the data subjects being able to show that they are aware of the fact that, and the extent to which, consent is given, when probed? 
 \longLabCriteriaOutcomeUse{\shortEffectiveness}{Accuracy}
}

\item[] \uSubCriteria{Is the level of awareness showed matching the intended degree of awareness?
 \longLabCriteriaOutcomeUse{\shortEffectiveness}{Completeness}
}

 \item[] \uSubCriteria{How sufficient are the safeguards taken to ensure that the data subjects are being aware of the fact that, and the extent to which, consent is given?
 \longLabCriteriaOutcomeUse{\shortEffectiveness}{Completeness}
}

 \item[] \uSubCriteria{\label{criteria_consent_declaration_TEFM}How much TEFM do the data subjects need to invest in order to become aware of the fact that, and the extent to which, consent is given?
 \longLabCriteriaOutcomeUse{\shortEfficiency}{\shortTEFM}
}

\vspace{1ex}
 \item[] To what degree do the data subjects perceive 
[\uSubCriteriaCount themselves as being aware, and
\uSubCriteriaCount that enough safeguards have been taken to help them become aware]
of the fact that, and the extent to which, consent is given?
\longLabCriteriaOutcomeUse{\shortSatisfaction}{Cognitive responses}
\end{itemize}
} 

\longPaper{
\uCriteria{\label{criteria_consent_intelligibleEasilyAccessible}Is the declaration of consent, pre-formulated by the controller, provided in an intelligible and easily accessible form, using clear and plain language, and not containing unfair terms?}
\shortLabCriteriaGoalBased{goal_consent_intelligible}\shortLabCriteriaType{Consent}
\begin{itemize}
 \item[] How much TEFM do the data subjects need to invest in order to 
[\uSubCriteriaCount\label{criteria_consent_intelligibleEasilyAccessible_TEFM_access} access, 
\uSubCriteriaCount\label{criteria_consent_intelligibleEasilyAccessible_TEFM_read} read, and
\uSubCriteriaCount\label{criteria_consent_intelligibleEasilyAccessible_TEFM_understand} understand]
the declaration of consent?
\shortLabCriteriaMeasure{\shortObjective}%
\longLabCriteriaOutcomeUse{\shortEfficiency}{\shortTEFM}
\vspace{1ex}

\vspace{1ex}
 \item[] How is the TEFM spent in relation to the TEFM expected by
[\uSubCriteriaCount the controllers, or
\uSubCriteriaCount the data subjects]?
Are the differences reasonable?\jjshort{Need to decide the exact labels for this one.}
\vspace{1ex}

\vspace{1ex}
 \item[] To what degree do the data subjects perceive 
[\uSubCriteriaCount the terms as unfair, 
\uSubCriteriaCount the language of the declaration of consent as clear and plain, and
\uSubCriteriaCount the declaration of consent as being intelligible an having an easily accessible form]?
\shortLabCriteriaMeasure{\shortPerceived}%
\longLabCriteriaOutcomeUse{\shortSatisfaction}{Cognitive responses}
\end{itemize}

\begin{remark}
 The criterion~\refuc{criteria_consent_intelligibleEasilyAccessible} is similar to the criteria~\refuc{criteria_info_conciseEasilyAccessible} and \refuc{criteria_info_clearPlainLanguage}, only that it refers to the declaration of consent (or terms of services), and thus we expect that besides the above subcriteria one would also employ subcriteria analogous to those in \refuc{criteria_info_conciseEasilyAccessible}.x and \refuc{criteria_info_clearPlainLanguage}.x.
\end{remark}
} 

\longPaper{
\uCriteria{\label{criteria_consent_Distinguishable}Is the request for consent presented in a manner clearly distinguishable from the other matters?}
\shortLabCriteriaGoalBased{goal_consent_requestClearlyDistinguishable}\shortLabCriteriaType{Consent}
\begin{itemize}
 \item[] What is the percentage of the data subjects able to 
[\uSubCriteriaCount understand that their consent is requested, and
\uSubCriteriaCount clearly distinguish the request for consent from the other matters]
when probed?
\shortLabCriteriaMeasure{\shortPerceived}%
 \longLabCriteriaOutcomeUse{\shortEffectiveness}{Accuracy}
\vspace{1ex}

 \item[] \uSubCriteria{\label{criteria_consent_Distinguishable_TEFM}How much TEFM do the data subjects need to invest in order to distinguish the request for consent from the other matters?
 \longLabCriteriaOutcomeUse{\shortEfficiency}{\shortTEFM}
}

 \item[] \uSubCriteria{To what degree do the data subjects perceive the request for consent as clearly distinguishable from other matters?
\shortLabCriteriaMeasure{\shortPerceived}%
\longLabCriteriaOutcomeUse{\shortSatisfaction}{Cognitive responses}
}
\end{itemize}

\begin{remark}
 The criterion~\refuc{criteria_consent_Distinguishable} is to some extent similar to the criterion~\refuc{criteria_info_visibleLegible} only that it talks about distinguishability of the declaration of consent (or terms of services), and thus one can expect more subcriteria similar to those from \refuc{criteria_info_visibleLegible} to be useful.
\end{remark}
} 

\longPaper{
\uCriteria{Do the data subjects have free and genuine choice in giving consent?}
\shortLabCriteriaGoalBased{goal_consent_genuine}\shortLabCriteriaType{Consent}
\begin{itemize}
 \item[] \uSubCriteria{To what degree do the data subjects perceive the choice of consenting as free and genuine?
\shortLabCriteriaMeasure{\shortPerceived}%
\longLabCriteriaOutcomeUse{\shortSatisfaction}{Cognitive responses}
}

 \item[] \uSubCriteria{Are the data subjects being offered any alternatives in case of not being able/not wanting to consent?
\shortLabCriteriaMeasure{\shortObjective}%
 \longLabCriteriaOutcomeUse{\shortEffectiveness}{Completeness}
}
\end{itemize}
} 

\longPaper{
\uCriteria{Are the data subjects being able to refuse or withdraw consent without detriment?}
\shortLabCriteriaGoalBased{goal_consent_withdraw}\shortLabCriteriaType{Consent}
\begin{itemize}
 \item[] \uSubCriteria{How much TEFM losses are there for the data subjects, related to withdrawing the consent?
 \longLabCriteriaOutcomeUse{\shortEfficiency}{\shortTEFM}
}

 \item[] \uSubCriteria{When evaluating the overall consequences for the data subjects, in case of withdrawing the consent, what is the degree of impact on the data subjects?
\shortLabCriteriaMeasure{\shortObjective}%
 \longLabCriteriaOutcomeUse{\shortEffectiveness}{Accuracy}
}

 \item[] \uSubCriteria{To what degree do the data subjects perceive that it is detrimental for them to refuse or withdraw consent?
\shortLabCriteriaMeasure{\shortPerceived}%
\longLabCriteriaOutcomeUse{\shortSatisfaction}{Cognitive responses}
}
\end{itemize}
} 

\longPaper{
\uCriteria{Is it as easy to withdraw consent as it is to give consent?}
\shortLabCriteriaGoalBased{goal_consent_easyWithdraw}\shortLabCriteriaType{Consent}
\begin{itemize}
 \item[] \uSubCriteria{How much TEFM do the data subjects spend to withdraw consent? Compare this to the TEFM needed to give consent (i.e., sum up results from \refuc{criteria_consent_electronic_TEFM}, \refuc{criteria_consent_declaration_TEFM}, \refuc{criteria_consent_intelligibleEasilyAccessible_TEFM_access}-\refuc{criteria_consent_intelligibleEasilyAccessible_TEFM_understand}, and \refuc{criteria_consent_Distinguishable_TEFM}).
 \longLabCriteriaOutcomeUse{\shortEfficiency}{\shortTEFM}
}

 \item[] \uSubCriteria{Do the data subjects perceive withdrawing of the consent similarly easy to giving consent?
\shortLabCriteriaMeasure{\shortPerceived}%
\longLabCriteriaOutcomeUse{\shortSatisfaction}{Cognitive responses}
}
\end{itemize}
} 

\longPaper{
\subsubsection{UP criteria related to data subject rights}\

\uCriteria{Are the rights of the data subjects, under Articles 15 to 22, (i.e., right of access, right to rectification, right to erasure, right to restriction of processing, right to data portability, right to object, and rights related to automated individual decision-making) facilitated?}
\shortLabCriteriaGoalBased{goal_rights_facilitate}\shortLabCriteriaType{Rights}
\begin{itemize}
 \item[] \uSubCriteria{How much TEFM do the data subjects spend in order to exercise their rights?
 \longLabCriteriaOutcomeUse{\shortEfficiency}{\shortTEFM}
}

 \item[] \uSubCriteria{How many of the rights under Articles 15 to 22 are facilitated; and to what degree?
 \longLabCriteriaOutcomeUse{\shortEffectiveness}{Completeness}
}

 \item[] \uSubCriteria{To what degree do the data subjects perceive that their rights are facilitated?
\shortLabCriteriaMeasure{\shortPerceived}%
\longLabCriteriaOutcomeUse{\shortSatisfaction}{Cognitive responses}
}

\item[] \uSubCriteria{What is the percentage of the data subjects that are able to exercise their rights with ease, when probed? 
 \longLabCriteriaOutcomeUse{\shortEffectiveness}{Accuracy}
}
\end{itemize}

\uCriteria{Are the data subjects being aware of how to exercise their rights in relation to processing of personal data?}
\shortLabCriteriaGoalBased{goal_rights_awareness}\shortLabCriteriaType{Rights}
\begin{itemize}
 \item[] \uSubCriteria{To what degree do the data subjects feel that they are aware of how to exercise their rights in relation to processing of personal data?
\shortLabCriteriaMeasure{\shortPerceived}%
\longLabCriteriaOutcomeUse{\shortSatisfaction}{Cognitive responses}
}

\item[] \uSubCriteria{What is the percentage of data subjects being able to explain which are the ways they could use to exercise their rights in relation to processing of personal data, when probed? 
 \longLabCriteriaOutcomeUse{\shortEffectiveness}{Accuracy}
}
\end{itemize}

\uCriteria{Do the data subjects have the right of access to personal data that has been collected concerning them, and can they exercise this right easily and at reasonable intervals, in order to be aware of, and verify, the lawfulness of the processing?}
\shortLabCriteriaGoalBased{goal_rights_easilyAccess}\shortLabCriteriaType{Rights}
\begin{itemize}
 \item[] \uSubCriteria{How much TEFM do the data subjects spend in order to access the personal data that has been collected concerning them?
 \longLabCriteriaOutcomeUse{\shortEfficiency}{\shortTEFM}
}

\vspace{1ex}
 \item[] To what degree do the data subjects perceive
 [%
 \uSubCriteriaCount accessing the personal data as easy, 
 \uSubCriteriaCount the intervals they are given access to the data as reasonable, and
 \uSubCriteriaCount themselves as being aware of the lawfulness of the processing
 ]?
\shortLabCriteriaMeasure{\shortPerceived}%
\longLabCriteriaOutcomeUse{\shortSatisfaction}{Cognitive responses}
\vspace{1ex}

\vspace{1ex}
\item[] What is the percentage of the data subjects 
[%
\uSubCriteriaCount being able to access the personal data as easy as intended,
\uSubCriteriaCount found to be aware of the lawfulness of the processing,
\uSubCriteriaCount that can verify the lawfulness of the processing],
when probed? 
 \longLabCriteriaOutcomeUse{\shortEffectiveness}{Accuracy}

\vspace{1ex}
 \item[] \uSubCriteria{How much of the personal data concerning them are the data subjects being able to access?
\shortLabCriteriaMeasure{\shortObjective}%
 \longLabCriteriaOutcomeUse{\shortEffectiveness}{Completeness}
 }

\end{itemize}

\uCriteria{Is the right to object explicitly brought to the attention of the data subjects and presented clearly and separately from any other information, at the latest at the time of the first communication with the data subjects?}
\shortLabCriteriaGoalBased{goal_rights_rightToObject}\shortLabCriteriaType{Rights}
\begin{itemize}
 \item[] \uSubCriteria{How much TEFM do the data subjects spend to find the information related to the right to object?
 \longLabCriteriaOutcomeUse{\shortEfficiency}{\shortTEFM}
}

\vspace{1ex}
 \item[] What is the percentage of the data subjects being able to 
[\uSubCriteriaCount separate the right to object from any other information, and
\uSubCriteriaCount exercise their right to object]
 -- when probed?
\longLabCriteriaOutcomeUse{\shortEffectiveness}{Accuracy}
\vspace{1ex}

\vspace{1ex}
 \item[] To what degree do the data subjects perceive 
[\uSubCriteriaCount the right to object as clearly presented, and
\uSubCriteriaCount the way the right to object has been brought to their attention as explicit]?
\shortLabCriteriaMeasure{\shortObjective}%
 \longLabCriteriaOutcomeUse{\shortEffectiveness}{Completeness}
\end{itemize}

\uCriteria{Are the data subjects being aware of risks, rules, safeguards and rights in relation to the processing of their personal data?}
\shortLabCriteriaGoalBased{goal_risk_awareness}\shortLabCriteriaType{Rights}
\begin{itemize}
 \item[] \uSubCriteria{How much TEFM do the data subjects spend in order to become aware of the risks, rules, safeguards and rights in relation to the processing of their personal data?
 \longLabCriteriaOutcomeUse{\shortEfficiency}{\shortTEFM}
}

\item[] \uSubCriteria{How accurately can the data subjects remember which are the risks, rules, safeguards and rights in relation to the processing of their personal data, when probed? 
\longLabCriteriaOutcomeUse{\shortEfficiency}{Cognitive responses}
}

\item[] \uSubCriteria{How many of the risks, rules, safeguards and rights in relation to the processing of their personal data are the data subjects being able to remember? 
 \longLabCriteriaOutcomeUse{\shortEffectiveness}{Completeness}
}

 \item[] \uSubCriteria{To what degree do the data subjects feel that they are aware of the risks, rules, safeguards and rights to their personal data?
\shortLabCriteriaMeasure{\shortPerceived}%
\longLabCriteriaOutcomeUse{\shortSatisfaction}{Cognitive responses}
}

 \item[] \uSubCriteria{What is the percentage of the data subjects being able to understand the implications of the risks, rules, safeguards and rights to their personal data?
\shortLabCriteriaMeasure{\shortPerceived}%
 \longLabCriteriaOutcomeUse{\shortEffectiveness}{Accuracy}
}
\end{itemize}
} 

\longPaper{
\subsubsection{UP criteria related to the purpose of processing} \

\uCriteria{Is the specific purpose for which personal data are processed explicit?}
\shortLabCriteriaGoalBased{goal_purpose_explicit}\shortLabCriteriaType{Purpose}
\begin{itemize}
\item[] \uSubCriteria{How accurately can the data subjects remember the purpose? 
\longLabCriteriaOutcomeUse{\shortEfficiency}{Cognitive responses}
}

\item[] \uSubCriteria{How many of the purposes can the data subjects remember correctly when several purposes are given? 
 \longLabCriteriaOutcomeUse{\shortEffectiveness}{Completeness}
}

 \item[] \uSubCriteria{What is the percentage of the data subjects being able to show that they know what is the purpose for which personal data are processed?
\shortLabCriteriaMeasure{\shortPerceived}%
 \longLabCriteriaOutcomeUse{\shortEffectiveness}{Accuracy}
}
\end{itemize}

\uCriteria{\label{personal_data_processing_necessity_criteria}Is the personal data adequate, relevant and limited to what is necessary for the purposes for which they are processed?}
\shortLabCriteriaGoalBased{goal_purpose_adequate}\shortLabCriteriaType{Purpose}
\begin{itemize}
 \item[] \uSubCriteria{To what degree do the data subjects feel that the processing of their personal data are adequate, relevant and limited to what is necessary for the given purposes?
\shortLabCriteriaMeasure{\shortPerceived}%
\longLabCriteriaOutcomeUse{\shortSatisfaction}{Cognitive responses}
}

\item[] \uSubCriteria{How many aspects do the data subjects point out to be inadequate, irrelevant and less or more than necessary?
 \longLabCriteriaOutcomeUse{\shortEffectiveness}{Completeness}
}
\end{itemize}

} 

\longPaper{
\subsubsection{UP criteria related to the legitimate interest of either the processors or the data subjects}\

\uCriteria{Is the existence of a legitimate interest of a controller carefully assessed, taking into consideration the reasonable expectations of the data subjects based on their relationship with the controller?}
\shortLabCriteriaGoalBased{goal_legitimateInterest_controller}\shortLabCriteriaType{Legitimate}
\begin{itemize}
 \item[] \uSubCriteria{How much TEFM do the data subjects spend to assess the legitimate interest of the controller?
 \longLabCriteriaOutcomeUse{\shortEfficiency}{\shortTEFM}
}

 \item[] \uSubCriteria{To what degree do the data subjects perceive the legitimate interest of the controller as carefully assessed?
\shortLabCriteriaMeasure{\shortPerceived}%
\longLabCriteriaOutcomeUse{\shortSatisfaction}{Cognitive responses}
}
\end{itemize}

\uCriteria{Has it been assessed whether the interests and fundamental rights of the data subjects could override the interest of the controllers where personal data are processed in circumstances where the data subjects do not reasonably expect further processing?}
\shortLabCriteriaGoalBased{goal_legitimateInterest_dataSubject}\shortLabCriteriaType{Legitimate}
\begin{itemize}
 \item[] \uSubCriteria{How much is known by the controllers about which are the circumstances where the data subjects do not reasonably expect further processing? How much of these knowledge have been confirmed by the data subjects?
 \longLabCriteriaOutcomeUse{\shortEffectiveness}{Completeness}
}

 \item[] \uSubCriteria{Do the data subjects and their controllers have a mutual agreement on what is considered to be reasonable further processing?
\longLabCriteriaOutcomeUse{\shortEffectiveness}{Accuracy}
}
\end{itemize}

} 

\section{Interactions between the three axes}\label{sec_axesIntersections}

Characteristic to the data legislation text is that it always refers to how principles and rights intersect and depend on each other. In this section, we give examples of such references found in the recitals of GDPR, relevant for some of the identified usability goals. The recitals, though not legally biding, are meant to provide more details to the GDPR's articles. 
%
The lawfulness, fairness, and transparency of processing principles, and the right to be informed appear to be closely interrelated, having also the highest occurrence of usability goals.

\begin{enumerate}

 \item The UP criterion \longPaper{\refuc{criteria_generic_control}}\onlyShortPaper{\cite[UPC.1]{theTR}} refers to the control the data subjects have over their data. The criterion can be related to the \textit{right to data portability}, through the Recital (68), where due to the aim of strengthening the control of the data subject, the ``data subject should also be allowed to receive personal data concerning him or her, which he or she has provided to a controller in a structured, commonly used, machine-readable and interoperable format, and to transmit it to another controller ...''. The same UP criterion can also be linked to \textit{data security principle} through the provision in the Recital (75) where the ``risk to the rights and freedoms of natural persons'' can result in data subjects being deprived of their rights and freedoms or prevented from exercising control over their personal data. The ``risk to the rights and freedoms of natural persons'' is also mentioned by the 
\cite[pp. 131, 134]{fra2018handbook} in the context of \textit{data security principle}.

 \item The UP criteria \refuc{criteria_info_conciseEasilyAccessible} and \refuc{criteria_info_clearPlainLanguage} are related to the \textit{transparency of processing principle}, which is referred to directly in the Recital (58), where the respective goals are extracted from -- ``The principle of transparency requires that any information and communication related to the processing of those personal data ...'' -- as well as to the \textit{principles of lawfulness and fairness}, which are also directly referred to in the Recital (39) -- ``Any processing of the personal data should be lawful and fair''.

 \item The \longPaper{goals}\onlyShortPaper{goal} \longPaper{\refgoal{goal_risk_awareness} and}
 \refgoal{goal_rights_awareness} \longPaper{relate}\onlyShortPaper{relates} to the \textit{fairness and transparency of processing principles}, and \longPaper{are}\onlyShortPaper{is} placed under these respective categories also by the 
 \cite[pp. 117, 120]{fra2018handbook}.

 \item The \longPaper{goals \refgoal{goal_purpose_explicit} and \refgoal{goal_info_clearPlainLanguage_Second} are}%
 \onlyShortPaper{goal \refgoal{goal_info_clearPlainLanguage_Second} is}
 mentioned in the context of the transparency principle, in the Recital (39), where the information to be given to the data subject relates to the purpose of processing. This connects \textit{the principle of transparency} with \longPaper{\textit{the principle of purpose limitation} in the case of \refgoal{goal_purpose_explicit} and}
 \textit{the principle of data minimization}\longPaper{ in the case of \refgoal{goal_info_clearPlainLanguage_Second}}. 

 \item \longPaper{The UP criterion 
 \refuc{personal_data_processing_necessity_criteria} is based on the goal \refgoal{goal_purpose_adequate} 
 extracted from the the Recital (39) of GDPR.}%
\onlyShortPaper{The UP criterion \cite[UPC.22]{theTR} about
``the personal data [being] adequate, relevant and limited to what is necessary
for the purposes for which they are processed'', is based on the \cite[UPG.10]{theTR}, extracted from the Recital (39) of GDPR.}
This criterion is mentioned in Recital (39) as one of the requirements for complying with \textit{the transparency principle}, while also referring to the purpose of processing. 
This connects the present criterion also with the \textit{principle of data minimization} and in addition with the \textit{data protection by design principle}.
 The link between the last two principles can also be seen in the EuroPriSe criteria catalog, where data minimization is the focus of the \cite[1.2.1 Data Protection by Design and by Default, p.18]{EuroPriSe2017}.
 


\end{enumerate}

\workInProgress{
\section{Use Cases for Validation}\label{sec_UseCases}

The model\jjshort{Write about validation and more about the use cases in particular.} we propose is exemplified through three use cases that make use of cyber-physical systems technologies (also known as Internet of Things -- IoT -- or ubiquitous computing \cite{weiser1993ubiquitous}): 
Assisted Living and Community Care System, 
Air Quality Monitoring for healthy indoor environments, 
and Diabetes App. 
Our approach is especially relevant for the IoT technologies \cite{weiser1993ubiquitous,perera2013context,stankovic2014research,sicari2015security} as the privacy protection is even more variable and context-dependent, given their nature characterized by: ubiquity (can be embedded everywhere in our environments or be attached to our bodies), invisibility (their small sizes make them easy to be hidden), sensing (accurately perceiving the attributes of our environments through temperature, light, noise sensors or audio and video recording), and memory amplification (every action, movement of ourselves and our environments can be continuously and unobtrusively recorded) \cite{langheinrich2001privacy}. These attributes make the IoT technologies able to generate granular and intimate data about people and everything or everyone in their surroundings, by that reducing privacy to zero. 

The model we propose is meant to identify and consider the specifics related to industry, technology and social-cultural context, remaining at the same time sufficiently generic to be applicable to all IT services, systems and products. The model is thus suitable to be used for context aware systems by eliciting meaningful information about the context where the systems are deployed and the environment (the physical space, other devices or people) they are interacting with.
}

\workInProgress{

\section{UP criteria trade-offs (Future work?)}

Achieving some privacy goals might imply renouncing to or lower the expectations for other goals. Which privacy goals are prioritized, contribute to business differentiation and support some type of users better then others. We thus add support for mapping between preferences/expectations and privacy goals.  This lays the basis for future work as creating tools for the producers to clearly show why their products should be preferred over the others. Furthermore, out of this initiative, we envision the apparition of services that will specialize on reviewing IT products based on their privacy characteristics.\jjSecondary{Review this sentence.}

} 

\vspace{-1ex}
\section{Conclusion and Further Work}\label{sec_Conclusion}\vspace{-1ex}

The benefits of the \cubeName\ model are multiple:
(i) emphasizing both the perspectives of data subjects and of controllers;
(ii) representing visually on the three variability axes the existing rights and principles criteria from EuroPriSe, together with our new UP criteria;
(iii) visualizing intersections between the three axes;
(iv) allowing ordering of the criteria on each axis.

The theory behind our usability evaluation of privacy is based on the well established standards ISO/IEC29100:2011 and ISO 9241-11:2018. We worked directly with the GDPR text, guided by \cite{fra2018handbook}, which also inspired our structuring of the EuroPriSe criteria into rights and principles.
Our HCI and usability perspective on privacy is influenced by the seminal works  \cite{adams1999users,good2003usability,ackerman2005privacy,karat2012privacy,karat2005usability,patrick2003human,Cranor:2018:SSI:3170427.3185061}. 

\vspace{1ex}
To build the UP Cube we have: 
\vspace{-1ex}
\begin{itemize}
 \item identified from the GDPR text \textit{30 UP goals},
 \item created \textit{24 UP criteria}, each with measurable subcriteria, and
 \item restructured the criteria of EuroPriSe, laid as the basis of the \cubeName.
\end{itemize}

\subsubsection*{Further Work.}
The \cubeName\ is meant as the groundwork for building a certification methodology, extending EuroPriSe to evaluate the usability of privacy.
The proposed UP criteria are designed to produce measurable evaluations, useful for generating privacy labels in order to guide stakeholders when choosing technological products, by representing and visualizing the different levels of privacy.
%
To achieve this larger goal, one needs to investigate which existing HCI methods for usability testing should be used for each of the UP criteria, and in what way.

One example of such a usability method for measuring the perceived usability of a system is the System Usability Scale (SUS) \cite{brooke1996sus}, a ten-item attitude Likert scale questionnaire. The standard \cite[Annex B: Usability measurements]{ISO9241-11:2018} also gives examples of methods that produce measurements relevant for our UP criteria, s.a.\ observing the user behavior to identify the actual usability problems, or asking the users to carry out tasks in a real or simulated context of use and measuring the outcomes. The experts can also run heuristic evaluations following design principles, theories and standards from the design and cognitive fields.
More concrete examples of HCI methods and how these could be used for privacy and security solutions can be found in \cite{karat2005usability}. 

Which methods are appropriate to use, the number of test persons, and other test related concerns, depend on contextual factors, s.a.\ the type of technology, users and  industry. Defining the required context is what our model offers support for. 
However, more work (e.g., providing guidelines and examples) is needed on how the context of use can be established.

HCI practices conduct user studies throughout the whole lifecycle of a product. These studies are run by the company itself, with the help of HCI (User Experience or Interaction Design) experts. For certification, the accredited data protection assessors would be using the results provided by the company to answer the UP criteria questions. In the cases of not enough or not reliable results, the assessors can recommend/require further testing. 
It would be valuable to have guidelines, e.g., in the form of a check-list, to help the assessors with establishing if the results from the company are reliable and sufficient. Recommendations for the businesses are useful as well, to guide how to conduct privacy related user testing, so that the results would be reliable later for certification. 

With the same goal of achieving a complete methodology that can be taken in use by the accreditation bodies, building on the present model, one could create a visual representation of the evaluation, i.e., a translation of the measurements of usability of privacy provided by the UP criteria into a visually appealing privacy label. 
This should serve as a vertically graded scale to differentiate a customer product from another. According to ISO 9241-11:2018, ``where usability is higher then expected, the system, product or service can have a competitive advantage (e.g. customer retention, or customers who are willing to pay a premium)''. The visuals will be thought to come in addition to the GDPR compliance seal and reflect the usability of the privacy implemented. The purpose will be the same as for the methodology, to help the businesses that have already achieved GDPR compliance to further differentiate themselves on the market. From the point of view of the user of the product, the visual scale would offer support for choosing the service or product that best respects her privacy expectations.

To further validate our \cubeName\ model and for exemplification, we are applying the UP criteria to \textit{three use cases} taken from pilots done in an ongoing European project called Secure COnnected Trustable Things (SCOTT):
\textit{(i)} Assisted Living and Community Care System, 
\textit{(ii)} Air Quality Monitoring for healthy indoor environments, and 
\textit{(iii)} Diabetes App. 
These are examples of IoT systems 
\cite{weiser1993ubiquitous,stankovic2014research,sicari2015security}  
for which our model is especially relevant, as the privacy protection is even more variable and context-dependent. IoT technologies, due to their nature (i.e., ubiquity, invisibility, and continuous sensing) \cite{langheinrich2001privacy}, are able to generate granular and intimate data about people and everything or everyone in their surroundings, by that reducing privacy to zero.

\section*{Acknowledgements}
We would like to thank the anonymous reviewers for helping improve the paper, as well as the members of the PriSec group at Karlstad University, for useful discussions and help with the present research while doing a secondment there.

%
%
%
\bibliographystyle{splncs04}
\bibliography{mybib}

\begin{thebibliography}{10}
\providecommand{\url}[1]{\texttt{#1}}
\providecommand{\urlprefix}{URL }
\providecommand{\doi}[1]{https://doi.org/#1}

\bibitem{ISO/IEC29100:2011}
{Information technology -- Security techniques -- Privacy framework.} Standard
  ISO/IEC 29100:2011 (2011)

\bibitem{EU2016GDPR}
{Regulation (EU) 2016/679 of the European Parliament and of the Council of 27
  April 2016 on the protection of natural persons with regard to the processing
  of personal data and on the free movement of such data, and repealing
  Directive 95/46/EC}. {Official Journal of the European Union}  \textbf{L
  119/1} (2016)

\bibitem{HouseOfLoardsReport2016}
{The House of Lords EU Committee, European Union Committee’s report on Online
  Platforms and the Digital Single Market}  (2016),
  \url{https://publications.parliament.uk/pa/ld201516/ldselect/ldeucom/129/12909.htm#_idTextAnchor235}

\bibitem{EuroPriSe2017}
{EuroPriSe Criteria for the certification of IT products and IT-based services
  -- v201701}. Tech. rep. (2017),
  \url{https://www.european-privacy-seal.eu/AppFile/GetFile/6a29f2ca-f918-4fdf-a1a8-7ec186b2e78a}

\bibitem{ISO9241-11:2018}
{Ergonomics of human-system interaction -- Part 11: Usability: Definitions and
  concepts}. Standard ISO 9241-11:2018 (2018)

\bibitem{ackerman2005privacy}
Ackerman, M.S., Mainwaring, S.D.: {Privacy Issues and Human-Computer
  Interaction}. In: Cranor, L., Garfinkel, S. (eds.) {Security and usability:
  designing secure systems that people can use}, pp. 381--399. {O'Reilly}
  (2005)

\bibitem{adams1999users}
Adams, A., Sasse, M.A.: {Users are not the enemy}. {Communications of the ACM}
  \textbf{42}(12),  41--46 (1999)

\bibitem{balboni2018controversies}
Balboni, P., Dragan, T.: Controversies and challenges of trustmarks: Lessons
  for privacy and data protection seals. In: {Privacy and Data Protection
  Seals}, pp. 83--111. Springer (2018)

\bibitem{brooke1996sus}
Brooke, J.: {SUS -- A quick and dirty usability scale}. Usability evaluation in
  industry  \textbf{189}(194), ~4--7 (1996)

\bibitem{cavoukian2018privacy}
Cavoukian, A., Chibba, M.: {Privacy seals in the USA, Europe, Japan, Canada and
  Australia}. In: {Privacy and Data Protection Seals}, pp. 59--82. Springer
  (2018)

\bibitem{cooper2004inmates}
Cooper, A.: {The Inmates Are Running the Asylum -- Why High-Tech Products Drive
  Us Crazy and How to Restore the Sanity}. {Sams Publishing} (2004)

\bibitem{Cranor:2018:SSI:3170427.3185061}
Cranor, L.F.: {SIGCHI Social Impact Award Talk -- Making Privacy and Security
  More Usable}. CHI EA '18, ACM (2018). \doi{10.1145/3170427.3185061}

\bibitem{Cranor2018Security}
Cranor, L.F., Garfinkel, S.: {Security and usability: designing secure systems
  that people can use}. {O'Reilly} (2005)

\bibitem{Dumas1999}
Dumas, J.S., Redish, J.C.: {A Practical Guide to Usability Testing}. Intellect
  Books, {Revised} edn. (1999)

\bibitem{edelman2011adverse}
Edelman, B.: Adverse selection in online "trust" certifications and search
  results. Electronic Commerce Research and Applications  \textbf{10}(1),
  17--25 (2011)

\bibitem{fra2018handbook}
{European Union Agency for Fundamental Rights}: {Handbook on European data
  protection law -- 2018 edition}. {Luxembourg: Publications Office of the
  European Union} (2018)

\bibitem{good2003usability}
Good, N.S., Krekelberg, A.: {Usability and privacy: a study of Kazaa P2P
  file-sharing}. In: {Proceedings of the SIGCHI conference on Human factors in
  computing systems}. pp. 137--144. {ACM} (2003)

\bibitem{hansen2018schleswig}
Hansen, M.: {The Schleswig-Holstein data protection seal}. In: {Privacy and
  Data Protection Seals}, pp. 35--48. Springer (2018)

\bibitem{iachello2007end}
Iachello, G., Hong, J.: {End-user Privacy in Human-Computer Interaction}.
  {Foundations and Trends in Human-Computer Interaction}  \textbf{1}(1),
  1--137 (2007)

\bibitem{Johansen2020}
Johansen, J., Fischer-H{\"u}bner, S.: {Making GDPR Usable: A Model to Support
  Usability Evaluations of Privacy}. In: Friedewald, M., {\"O}nen, M., Lievens,
  E., Krenn, S., Fricker, S. (eds.) Privacy and Identity Management. Data for
  Better Living: AI and Privacy, IFIP Advances in Information and Communication
  Technology, vol.~576, pp. 275--291. Springer International Publishing (2020).
  \doi{10.1007/978-3-030-42504-3\_18}

\bibitem{kamara2018data}
Kamara, I., De~Hert, P.: {Data protection certification in the EU:
  Possibilities, Actors and Building Blocks in a reformed landscape}. In:
  {Privacy and Data Protection Seals}, pp. 7--34. Springer (2018)

\bibitem{karat2005usability}
Karat, C.M., Brodie, C., Karat, J.: {Usability design and evaluation for
  privacy and security solutions}. In: Cranor, L., Garfinkel, S. (eds.)
  {Security and usability: designing secure systems that people can use}, pp.
  47--74. {O'Reilly} (2005)

\bibitem{karat2012privacy}
Karat, C.M., Karat, J., Brodie, C.: {Privacy Security and Trust: Human-Computer
  Interaction Challenges and Opportunities at their Intersection}. {The
  Human-Computer Interaction Handbook} pp. 669--700 (2012)

\bibitem{langheinrich2001privacy}
Langheinrich, M.: {Privacy by design -- Principles of privacy-aware ubiquitous
  systems}. In: {International Conference on Ubiquitous Computing}. pp.
  273--291. Springer (2001)

\bibitem{nissim2017bridging}
Nissim, K., Bembenek, A., Wood, A., Bun, M., Gaboardi, M., Gasser, U., O'Brien,
  D.R., Steinke, T., Vadhan, S.: Bridging the gap between computer science and
  legal approaches to privacy. {Harvard Journal of Law \& Technology}
  \textbf{31}(2), ~687 (Spring 2018)

\bibitem{papakonstantinou2018introduction}
Papakonstantinou, V.: {Introduction: Privacy and Data Protection Seals}. In:
  {Privacy and Data Protection Seals}, pp.~1--6. Springer (2018)

\bibitem{patrick2003privacy}
Patrick, A.S., Kenny, S.: {From privacy legislation to interface design:
  Implementing information privacy in human-computer interactions}. In:
  {International Workshop on Privacy Enhancing Technologies}. pp. 107--124.
  Springer (2003)

\bibitem{patrick2003human}
Patrick, A.S., Kenny, S., Holmes, C., van Breukelen, M.: {Human Computer
  Interaction}. In: {Handbook for Privacy and Privacy-Enhancing Technologies:
  The case of Intelligent Software Agents}, chap.~12, pp. 249--290 (2003)

\bibitem{preece2015interaction}
Preece, J., Rogers, Y., Sharp, H.: {Interaction design: beyond human-computer
  interaction}. {John Wiley \& Sons} (2015)

\bibitem{schneier2015data}
Schneier, B.: {Data and Goliath: The hidden battles to collect your data and
  control your world}. {WW Norton \& Company} (2015)

\bibitem{sicari2015security}
Sicari, S., Rizzardi, A., Grieco, L.A., Coen-Porisini, A.: {Security, privacy
  and trust in Internet of Things: The road ahead}. Computer networks
  \textbf{76},  146--164 (2015)

\bibitem{stankovic2014research}
Stankovic, J.A.: Research directions for the internet of things. IEEE Internet
  of Things Journal  \textbf{1}(1), ~3--9 (2014)

\bibitem{weiser1993ubiquitous}
Weiser, M.: Ubiquitous computing. Computer (10),  71--72 (1993)

\bibitem{whitten1999johnny}
Whitten, A., Tygar, J.D.: {Why Johnny Can't Encrypt: A Usability Evaluation of
  PGP 5.0.} In: {USENIX Security Symposium}. vol.~348 (1999)

\end{thebibliography}
%
%
%
%
%
%
%

\longPaper{
\newpage
\section{Annexes}
\subsection{Annex A: A list of the Recitals and Articles of GDPR, in full text, from which the usability goals have been extracted.}

(6) Rapid technological developments and globalisation have brought new challenges for the protection of personal data. The scale of the collection and sharing of personal data has increased significantly. Technology allows both private companies and public authorities to make use of personal data on an unprecedented scale in order to pursue their activities. Natural persons increasingly make personal information available publicly and globally. Technology has transformed both the economy and social life, and should further facilitate the free flow of personal data within the Union and the transfer to third countries and international organisations, while ensuring a high level of the protection of personal data.

(7) Those developments require a strong and more coherent data protection framework in the Union, backed by strong enforcement, given the importance of creating the trust that will allow the digital economy to develop across the internal market. Natural persons should have control of their own personal data. Legal and practical certainty for natural persons, economic operators and public authorities should be enhanced.

(32) Consent should be given by a clear affirmative act establishing a freely given, specific, informed and unambiguous indication of the data subject's agreement to the processing of personal data relating to him or her, such as by a written statement, including by electronic means, or an oral statement. This could include ticking a box when visiting an internet website, choosing technical settings for information society services or another statement or conduct which clearly indicates in this context the data subject's acceptance of the proposed processing of his or her personal data. Silence, pre-ticked boxes or inactivity should not therefore constitute consent. Consent should cover all processing activities carried out for the same purpose or purposes. When the processing has multiple purposes, consent should be given for all of them. If the data subject's consent is to be
given following a request by electronic means, the request must be clear, concise and not unnecessarily disruptive to the use of the service for which it is provided.

(39)  Any processing of personal data should be lawful and fair. It should be transparent to natural persons that personal data concerning them are collected, used, consulted or otherwise processed and to what extent the personal data are or will be processed. The principle of transparency requires that any information and communi­cation relating to the processing of those personal data be easily accessible and easy to understand, and that clear and plain language be used. That principle concerns, in particular, information to the data subjects on the identity of the controller and the purposes of the processing and further information to ensure fair and transparent processing in respect of the natural persons concerned and their right to obtain confirmation and communication of personal data concerning them which are being processed. Natural persons should be made
aware of risks, rules, safeguards and rights in relation to the processing of personal data and how to exercise their rights in relation to such processing. In particular, the specific purposes for which personal data are processed should be explicit and legitimate and determined at the time of the collection of the personal data. The personal data should be adequate, relevant and limited to what is necessary for the purposes for which they are processed. This requires, in particular, ensuring that the period for which the personal data are stored is limited to a strict minimum. Personal data should be processed only if the purpose of the processing could not reasonably be fulfilled by other means. In order to ensure that the personal data are not kept longer than necessary, time limits
should be established by the controller for erasure or for a periodic review. Every reasonable step should be taken to ensure that personal data which are inaccurate are rectified or deleted. Personal data should be processed in a manner that ensures appropriate security and confidentiality of the personal data, including for preventing unauthorised access to or use of personal data and the equipment used for the processing.

(42) Where processing is based on the data subject's consent, the controller should be able to demonstrate that the data subject has given consent to the processing operation. In particular in the context of a written declaration on another matter, safeguards should ensure that the data subject is aware of the fact that and the extent to which consent is given. In accordance with Council Directive 93/13/EEC (1) a declaration of consent pre-formulated by the controller should be provided in an intelligible and easily accessible form, using clear and plain language and it should not contain unfair terms. For consent to be informed, the data subject should be aware at
least of the identity of the controller and the purposes of the processing for which the personal data are intended. Consent should not be regarded as freely given if the data subject has no genuine or free choice or is unable to refuse or withdraw consent without detriment.

(43) In order to ensure that consent is freely given, consent should not provide a valid legal ground for the processing of personal data in a specific case where there is a clear imbalance between the data subject and the controller, in particular where the controller is a public authority and it is therefore unlikely that consent was freely given in all the circumstances of that specific situation. Consent is presumed not to be freely given if it does not allow separate consent to be given to different personal data processing operations despite it being appropriate in the individual case, or if the performance of a contract, including the provision of a service, is dependent on the consent despite such consent not being necessary for such performance.

(47)  The legitimate interests of a controller, including those of a controller to which the personal data may be disclosed, or of a third party, may provide a legal basis for processing, provided that the interests or the fundamental rights and freedoms of the data subject are not overriding, taking into consideration the reasonable expectations of data subjects based on their relationship with the controller. Such legitimate interest could exist for example where there is a relevant and appropriate relationship between the data subject and the controller in situations such as where the data subject is a client or in the service of the controller. At any rate the existence of a legitimate interest would need careful assessment including whether a data subject can reasonably expect at the time and in the context of the collection of the personal data that processing for that purpose may take place.
The interests and fundamental rights of the data subject could in particular override the interest of the data controller where personal data are processed in circumstances where data subjects do not reasonably expect further processing. Given that it is for the legislator to provide by law for the legal basis for public authorities to process personal data, that legal basis should not apply to the processing by public authorities in the performance of their tasks. The processing of personal data strictly necessary for the purposes of preventing fraud also constitutes a legitimate interest of the data controller concerned. The processing of personal data for direct marketing purposes may be regarded as carried out for a legitimate interest.

(58) The principle of transparency requires that any information addressed to the public or to the data subject be concise, easily accessible and easy to understand, and that clear and plain language and, additionally, where appropriate, visualisation be used. Such information could be provided in electronic form, for example, when addressed to the public, through a website. This is of particular relevance in situations where the proliferation of actors and the technological complexity of practice make it difficult for the data subject to know and understand whether, by whom and for what purpose personal data relating to him or her are being collected, such as in the case of online advertising. Given that children merit specific protection, any information and communication,
where processing is addressed to a child, should be in such a clear and plain language that the child can easily understand.

(60) The principles of fair and transparent processing require that the data subject be informed of the existence of the processing operation and its purposes. The controller should provide the data subject with any further information necessary to ensure fair and transparent processing taking into account the specific circumstances and context in which the personal data are processed. Furthermore, the data subject should be informed of the existence of profiling and the consequences of such profiling. Where the personal data are collected from the data subject, the data subject should also be informed whether he or she is obliged to provide the personal data and of the consequences, where he or she does not provide such data. That information may be provided in
combination with standardised icons in order to give in an easily visible, intelligible and clearly legible manner, a meaningful overview of the intended processing. Where the icons are presented electronically, they should be machine-readable.

(63)  A data subject should have the right of access to personal data which have been collected concerning him or her, and to exercise that right easily and at reasonable intervals, in order to be aware of, and verify, the lawfulness of the processing. This includes the right for data subjects to have access to data concerning their health, for example the data in their medical records containing information such as diagnoses, examination results, assessments by treating physicians and any treatment or interventions provided. Every data subject should therefore have the right to know and obtain communication in particular with regard to the purposes for which the personal data are processed, where possible the period for which the personal data are processed, the recipients of the personal data, the logic involved in any automatic personal data processing and, at least when
based on profiling, the consequences of such processing. Where possible, the controller should be able to provide remote access to a secure system which would provide the data subject with direct access to his or her personal data. That right should not adversely affect the rights or freedoms of others, including trade secrets or intellectual property and in particular the copyright protecting the software. However, the result of those considerations should not be a refusal to provide all information to the data subject. Where the controller processes a large quantity of information concerning the data subject, the controller should be able to request that, before the information is delivered, the data subject specify the information or processing activities to which the request
relates.

(100) In order to enhance transparency and compliance with this Regulation, the establishment of certification mechanisms and data protection seals and marks should be encouraged, allowing data subjects to quickly assess the level of data protection of relevant products and services.

\textit{CHAPTER II. Principles}

\textit{Article 7. Conditions for consent}

2.  If the data subject's consent is given in the context of a written declaration which also concerns other matters, the request for consent shall be presented in a manner which is clearly distinguishable from the other matters, in an intelligible and easily accessible form, using clear and plain language. Any part of such a declaration which constitutes an infringement of this Regulation shall not be binding.

3.  The data subject shall have the right to withdraw his or her consent at any time. The withdrawal of consent shall not affect the lawfulness of processing based on consent before its withdrawal. Prior to giving consent, the data subject shall be informed thereof. It shall be as easy to withdraw as to give consent.

\textit{CHAPTER III. Rights of the data subject}

\textit{Section 1. Transparency and modalities}

\textit{Article 12. Transparent information, communication and modalities for the exercise of the rights of the data subject}

1.  The controller shall take appropriate measures to provide any information referred to in Articles 13 and 14 and any communication under Articles 15 to 22 and 34 relating to processing to the data subject in a concise, transparent, intelligible and easily accessible form, using clear and plain language, in particular for any information addressed specifically to a child. The information shall be provided in writing, or by other means, including, where appropriate, by electronic means. When requested by the data subject, the information may be provided orally, provided that the identity of the data subject is proven by other means.

2.  The controller shall facilitate the exercise of data subject rights under Articles 15 to 22. In the cases referred to in Article 11(2), the controller shall not refuse to act on the request of the data subject for exercising his or her rights under Articles 15 to 22, unless the controller demonstrates that it is not in a position to identify the data subject.

7.  The information to be provided to data subjects pursuant to Articles 13 and 14 may be provided in combination with standardised icons in order to give in an easily visible, intelligible and clearly legible manner a meaningful overview of the intended processing. Where the icons are presented electronically they shall be machine-readable.

\textit{Section 2. Information and access to personal data}

\textit{Article 15. Right of access by the data subject}

1. The data subject shall have the right to obtain from the controller confirmation as to whether or not personal data concerning him or her are being processed, and, where that is the case, access to the personal data and the following information:
(h) the existence of automated decision-making, including profiling, referred to in Article 22(1) and (4) and, at least in those cases, meaningful information about the logic involved, as well as the significance and the envisaged consequences of such processing for the data subject.

\textit{Section 3. Rectification and erasure}

\textit{Article 17. Right to erasure ('right to be forgotten')}

2.  Where the controller has made the personal data public and is obliged pursuant to paragraph 1 to erase the personal data, the controller, taking account of available technology and the cost of implementation, shall take reasonable steps, including technical measures, to inform controllers which are processing the personal data that the data subject has requested the erasure by such controllers of any links to, or copy or replication of, those personal data.

\textit{Section 4. Right to object and automated individual decision-making}

\textit{Article 21. Right to object}

4.  At the latest at the time of the first communication with the data subject, the right referred to in paragraphs 1 and 2 shall be explicitly brought to the attention of the data subject and shall be presented clearly and separately from any other information.

\textit{Article 22. Automated individual decision-making, including profiling}

1.  The data subject shall have the right not to be subject to a decision based solely on automated processing, including profiling, which produces legal effects concerning him or her or similarly significantly affects him or her.
}  

\end{document}